\documentclass[letter]{aa} 
\pdfoutput=1 
\usepackage{amssymb} 
\usepackage{txfonts}
\usepackage{appendix} 
\usepackage{soul} 
\usepackage[colorlinks=true,
  linkcolor=blue, filecolor=magenta, citecolor=blue,
  urlcolor=cyan]{hyperref}

\newcommand{\xmm}{{\textit{XMM-Newton}}\/}
\newcommand{\swift}{{\textit{Swift}}\/}
\newcommand{\cha}{{\textit{Chandra}}\/}

\newcommand\T{\rule{0pt}{2.4ex}}       
\newcommand\B{\rule[-1.0ex]{0pt}{0pt}} 

\begin{document}

\title{Alive and kicking: A new QPE phase in GSN~069 revealing a
  quiescent luminosity threshold for QPEs}

   \author{G. Miniutti\inst{1}\fnmsep\thanks{gminiutti@cab.inta-csic.es}
   \and M. Giustini\inst{1} 
   \and R. Arcodia\inst{2}\fnmsep\thanks{Einstein Fellow} 
   \and R. D. Saxton\inst{3} 
   \and J. Chakraborty\inst{2} 
   \and A. M. Read\inst{4} 
   \and E. Kara\inst{2}
}

   \institute{Centro de Astrobiolog\'ia (CAB), CSIC-INTA, Camino Bajo
     del Castillo s/n, ESAC campus, 28692 Villanueva de la Ca\~nada,
     Madrid, Spain 
   \and MIT Kavli Institute for Astrophysics and Space Research,
   Massachusetts Institute of Technology, Cambridge, MA 02139   
   \and Telespazio-Vega UK for ESA, Operations Department, European
   Space Astronomy Centre (ESAC), Villanueva de la Ca\~nada, E-28692
   Madrid, Spain
   \and Department of Physics \& Astronomy, University of Leicester,
   Leicester, LE1 7RH, UK }

   \date{Received  / Accepted  }

\abstract {X-ray quasi-periodic eruptions (QPEs) are intense repeating
  soft X-ray bursts from the nuclei of nearby galaxies. Their physical
  origin is still largely unconstrained, and several theoretical
  models have been proposed ranging from disc instabilities to impacts
  between an orbiting companion and the existing accretion disc around
  the primary, or episodic mass transfer at pericentre in an extreme
  mass-ratio binary. We present here results from a recent
  \xmm\ observation of GSN~069, the galactic nucleus where QPEs were
  first discovered. After about two years of absence, QPEs have
  reappeared in GSN~069, and we detect two consecutive QPEs separated
  by a much shorter recurrence time than ever before. Moreover, their
  intensities and peak temperatures are remarkably different, a novel
  addition to the QPE phenomenology. We study the QPE spectral
  properties from all \xmm\ observations assuming QPEs to either
  represent an additional emission component superimposed on that from
  the disc, or the transient evolution of the disc emission itself. In
  the former scenario, QPEs are consistent with black-body emission
  from a region that expands by a factor of $2$-$3$ during the
  individual QPE evolution with radius $\simeq 5$-$10\times
  10^{10}$~cm at QPE peak. In the alternative non-additive scenario,
  QPEs originate from a region with an area $\sim 6$-$30$ times
  smaller than the quiescent state X-ray emission, with the smallest
  regions corresponding to the hottest and most luminous
  eruptions. The QPE reappearance reveals that eruptions are only
  present below a quiescent luminosity threshold corresponding to an
  Eddington ratio $\lambda_{\rm thresh} \simeq 0.4\pm 0.2$ for a
  $10^6$~M$_\sun$ black hole. The disappearance of QPEs above
  $\lambda_{\rm thresh}$ is most likely driven by the ratio of QPE to
  quiescence temperatures, $kT_{\rm QPE}/kT_{\rm quiesc}$, approaching
  unity at high quiescent luminosity, making QPE detection
  challenging, if not impossible, above threshold. We briefly discuss
  some of the consequences of our results on the proposed models for
  the QPE physical origin.}

\keywords{Galaxies: nuclei --- Galaxies: individual: GSN~069 ---
  Accretion, accretion disks --- Black Hole Physics --- X-rays:
  individuals: GSN~069}

\titlerunning{A new QPE phase in GSN~069}
\authorrunning{Miniutti et al.}

\maketitle

\section{Introduction}
\label{sec:intro}

First discovered in the nucleus of the galaxy GSN~069
\citep{2019Natur.573..381M}, X-ray quasi-periodic eruptions (QPEs) are
one of the most recent examples of extreme X-ray variability
associated with supermassive black holes (SMBHs). QPEs are fast
intense soft X-ray bursts repeating every few hours that stand out
with respect to an otherwise stable quiescent X-ray emission, likely
from an existing accretion disc. Following their first detection in
GSN~069, QPEs have been identified in the nuclei of other four
galaxies to date: RX~J1301.9+2747 \citep{2020A&A...636L...2G},
eRO-QPE1 and eRO-QPE2 \citep{2021Natur.592..704A,2022A&A...662A..49A},
and XMMSL1~J024916.6-04124 \citep{2021ApJ...921L..40C}. A further QPE
candidate is 2XMM~J123103.2+110648 \citep{2012ApJ...752..154T},
although its X-ray variability is more reminiscent of a quasi-periodic
oscillation (QPO) than of QPEs
\citep{2013ApJ...776L..10L,2023MNRAS.518.3428W}, and the source has
been proposed to represent a descendant of QPE sources
\citep{2023MNRAS.tmpL..51K}.

Quasi-periodic eruptions have thermal-like X-ray spectra with
temperatures evolving from $kT\simeq 50$-$80$~eV to $\simeq
100$-$250$~eV and back in about one to few hours with duty cycle
(duration over recurrence) of $10$-$30$\% and peak X-ray luminosity
of $10^{42}$-$10^{43}$~erg~s$^{-1}$, depending on the specific
source. QPE host galaxies harbour SMBHs of relatively low mass
($\simeq 10^5$-$10^7$~M$_\sun$ at most), and they are best classified
as post-starburst galaxies \citep{2022A&A...659L...2W}, a population
that is similar to the preferred tidal disruption events (TDEs) hosts,
as is the low SMBH mass
\citep{2014ApJ...793...38A,2020SSRv..216...32F}. Two out of the five
currently known QPE sources, GSN~069 and XMMSL1~J024916.6-04124, have
been directly associated with X-ray TDEs
\citep{2018ApJ...857L..16S,2021ApJ...920L..25S,2023A&A...670A..93M,2021ApJ...921L..40C}. A
further QPE candidate was found during the X-ray decay of an optically
detected TDE, strengthening the possible connection between QPEs and
TDEs \citep{tormund}. Despite being unobscured in the X-rays, optical
and UV spectra never show signs of broad emission lines (BELs)
\citep[see][]{2022A&A...659L...2W}. The lack of BELs indicates
either that an active nucleus switched off leaving only relic narrow
lines, or that the accretion flow is unable to support a mature broad
line region, perhaps being too compact, as expected for instance in
TDEs.

We focus here on the discovery-source, GSN~069 \citep[see
  also][]{2013MNRAS.433.1764M}, whose long-term evolution over the
past $\sim 12$~yr has been discussed by \citet{2023A&A...670A..93M}
together with the main properties of its QPEs. The evolution has so
far been consistent with two repeating, possibly partial X-ray TDEs
about $9$~yr apart (hereafter TDE~1 and TDE~2), although it is
difficult to exclude a different origin for the observed long-term
X-ray variability as the luminosity has decayed by only a factor of
a few in $\sim 9$~yr. A set of 15 QPEs with well-defined properties were
detected in four observations performed towards the end of the decay
of TDE~1 (XMM3 to XMM5 in Table~\ref{tab:obs}, where all high-quality
observations of GSN~069 are reported). These QPEs (hereafter referred
to as regular QPEs) were separated by $\simeq 32$~ks on average. QPE
intensities alternated, and stronger (weaker) QPEs were systematically
followed by longer (shorter) recurrence times with $T_{\rm long}
\simeq 33$~ks ($T_{\rm short} \simeq 31$~ks) on average. Moreover, the
higher (lower) the intensity ratio between consecutive QPEs, the
longer (shorter) the recurrence time between them. A further set of
four weaker QPEs was detected in an \xmm\ observation performed during
the rise of TDE~2 (XMM6 in Table~\ref{tab:obs}), when the quiescent
level was significantly brighter than in any previous observation with
QPEs. This set of weaker QPEs (hereafter referred to as irregular
QPEs) broke the previous regular patterns and, while intensities still
roughly alternated, recurrence times did not, signalling that the
system was undergoing  major change at that epoch (possibly
consistent with XMM6 being performed during the rise of TDE~2). During
all \xmm\ observations with regular QPEs, a lower intensity QPO of the
otherwise stable quiescent emission was also detected with period
equal to the average recurrence time between consecutive QPEs. Such
QPO-like variability was absent during the XMM6 observation with
weaker irregular QPEs. Eruptions have since then disappeared, as
demonstrated by long-enough exposures from May 2020 to December 2021
(XMM7 to XMM11 in Table~\ref{tab:obs}).

We report results from a new \xmm\ observation of GSN~069 performed on
7 July 2022, about seven months after the last observation where no
QPEs were detected. As the overall X-ray flux was decaying after
TDE~2, searching for the reappearance of QPEs at flux levels similar
to the previous regular QPE phase was the main motivation for
observing GSN~069 with \xmm. We used the same data reduction
procedures of \citet{2023A&A...670A..93M} for all X-ray observations
and we refer to their work for details.

\section{A new QPE phase}
\label{sec:newQPEs}

\begin{figure}
\centering \includegraphics[width=0.96\columnwidth]{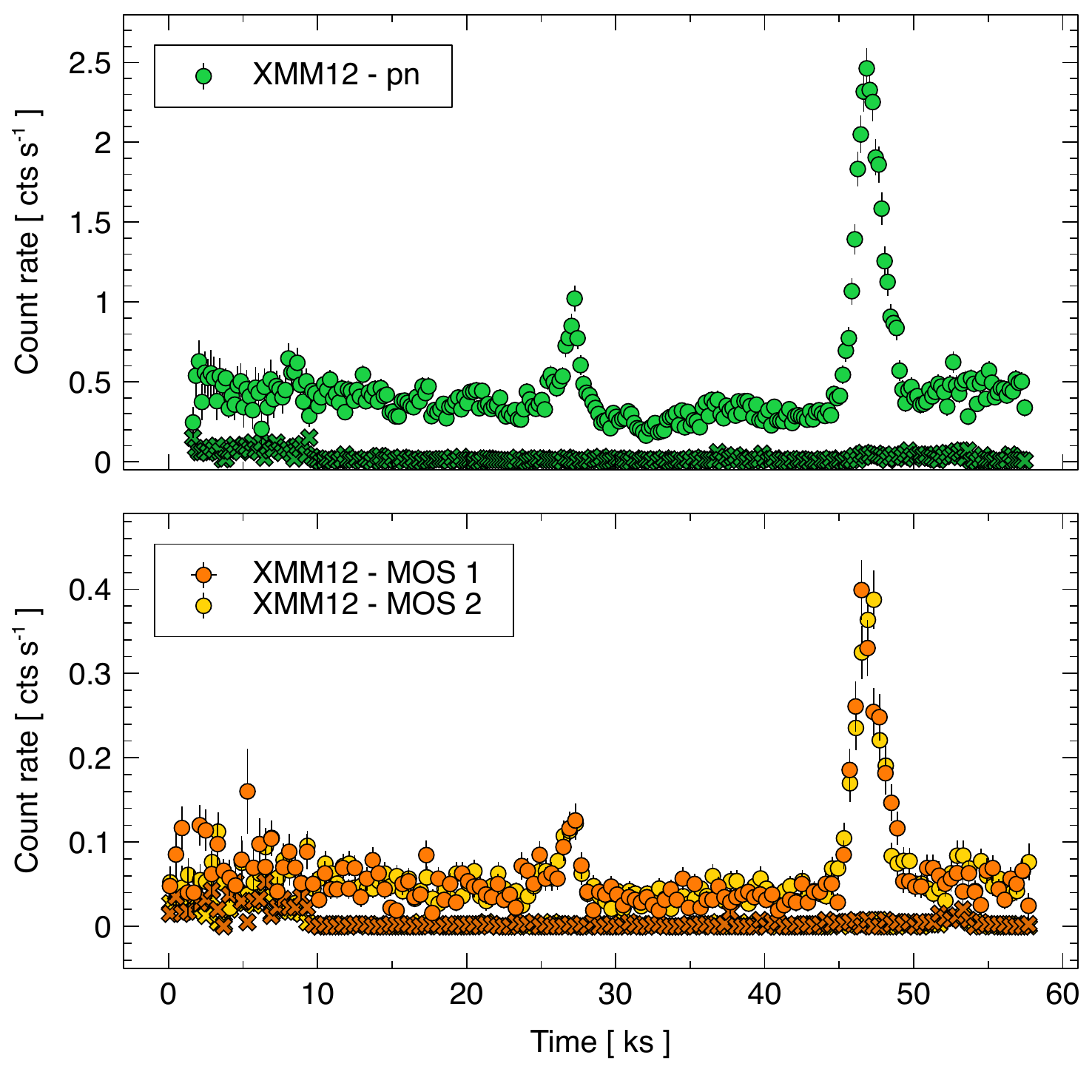}
{\vspace{-0.2cm}}
\caption{EPIC light curves from the XMM12 observation. Light
  curves in the $0.2$-$1$~keV with time bins of $200$~s (pn) and
  $400$~s (MOS) are shown. Background light curves, rescaled
  to the same extraction area, are also shown to highlight the slightly higher
  background during the initial $\sim 10$~ks of the EPIC exposures. The
  start of the MOS~1 exposure is taken as origin for the time-axis in all
  cases.}
\label{fig:XMM12lc}
\end{figure}

The EPIC pn, MOS1, and MOS2 $0.2$-$1$~keV light curves during the new
\xmm\ observation (hereafter XMM12) are shown in Fig.~\ref{fig:XMM12lc} and
demonstrate that relatively high-amplitude QPEs have reappeared in
GSN~069 after $2.0\pm 0.5$~yr of absence. As done in previous QPE
analyses \citep{2023A&A...670A..93M}, we fitted the EPIC pn light
curves with a baseline phenomenological model comprising a constant,
representing the quiescent level emission, and two Gaussian functions
describing the QPEs. The fit was performed in both the $0.2$-$1$~keV
and the $0.4$-$1$~keV bands to ease comparison with previous work
where the restricted energy band was used to include three QPEs from a
\cha\ observation. Results are reported in Table~\ref{tab:baseline}.

We detect one weak and one strong QPE during XMM12, separated by
$\simeq 20$~ks. The separation is significantly shorter than the
typical recurrence time during the previous regular QPE phase ($\simeq
32$~ks), and even than the shortest recurrence time ever detected
during the irregular XMM6 observation ($\simeq 26$~ks). No QPEs are
detected in the first $\simeq 27$~ks of the EPIC exposures. QPEs are
therefore no longer strictly quasi-periodic or, at least, the
difference between long and short recurrence times has significantly
increased with respect to the previous QPE phase. The intensity ratio
between the weak and strong QPE in XMM12 ($\simeq 0.26$ in the
$0.2$-$1$~keV band) is also much lower than between any consecutive
QPE pair in the previous regular phase ($\gtrsim 0.67$).

\begin{figure}
\centering
\includegraphics[width=0.96\columnwidth]{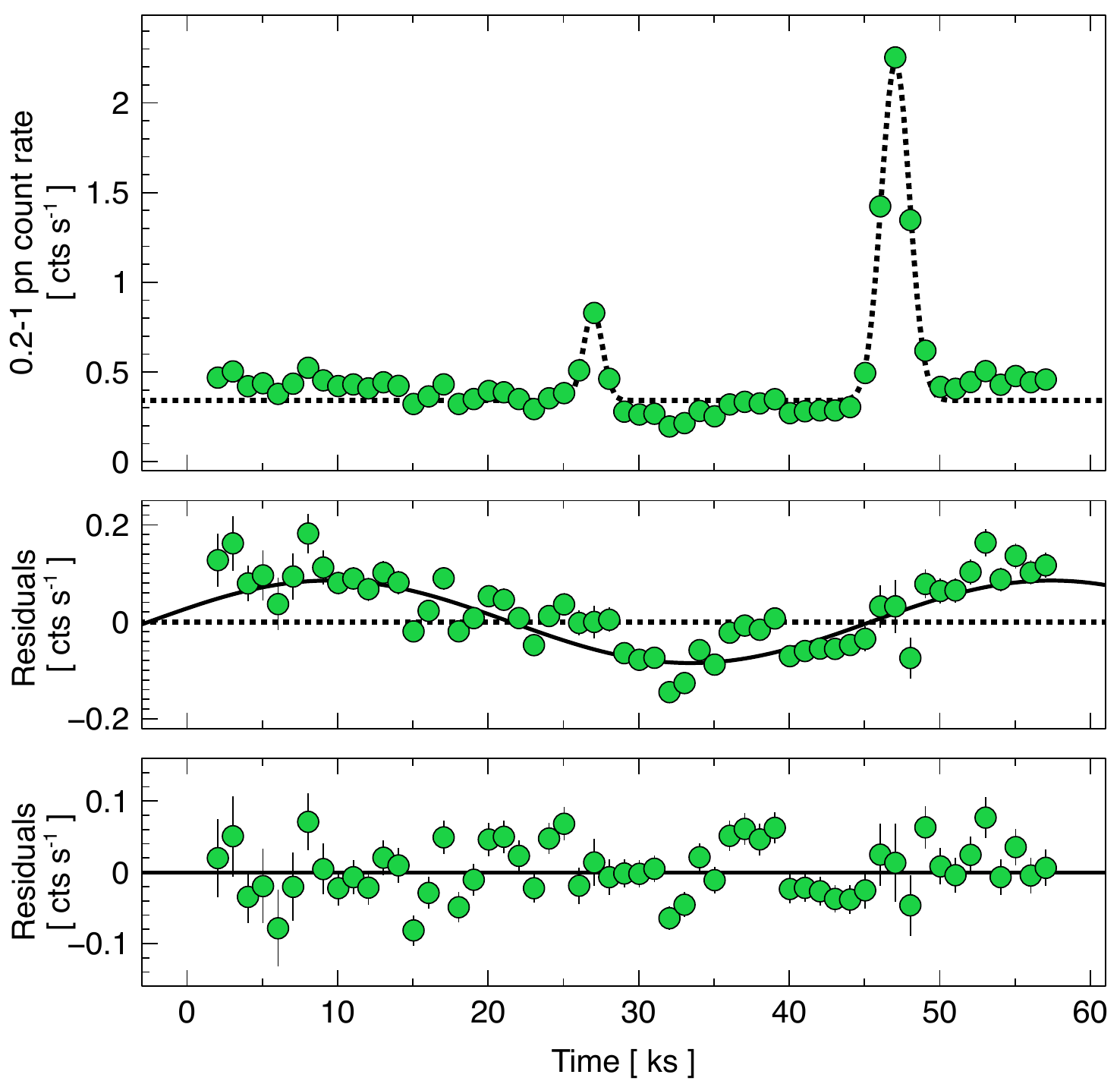}
\caption{Quiescent level QPO. Upper panel: $0.2$-$1$~keV pn light
  curve with time bins of $1\,000$~s, together with its baseline
  best-fitting model. Middle panel: Resulting residuals
  and, as a solid line, a sine function that is plotted to guide the
  eye. Lower panel: Residuals once the original light
  curve is fitted by adding a sine function to the baseline model.}
\label{fig:XMM12fits}
\end{figure}

As is clear in Fig.~\ref{fig:XMM12lc}, the quiescent level during the
XMM12 observation is not constant during the exposure, but exhibits
variability on timescales of a few tens of ks. In order to study it in
more detail, we rebinned the original light curve by a factor of five
to increase the signal-to-noise ratio. In the upper panel of
Fig.~\ref{fig:XMM12fits}, we show the rebinned EPIC pn light curve
together with the best-fitting baseline model resulting in $\chi^2 =
550$ for $49$ degrees of freedom. While the model describes QPEs well,
the overall statistical quality of the fit is very poor. Residuals for
the baseline model fit are shown in the middle panel of
Fig.~\ref{fig:XMM12fits}. The shape of the residuals suggest a
possible sinusoidal modulation, or QPO, as indicated by the solid line
(plotted to guide the eye). By refitting the
original light curve (upper panel) with the addition of a sine
function, the statistical quality of the fit improves by $\Delta\chi^2
= -410$ for three degrees of freedom, and the resulting residual light
curve is shown in the lower panel of Fig.~\ref{fig:XMM12fits}. We
derive a best-fitting period $P = 54\pm 4$~ks, and the results are reported
in Table~\ref{tab:sine}.

As only one cycle is detected and, furthermore, with a period
suspiciously similar to the exposure duration, the observed QPO
candidate cannot be considered significant from a statistical point of
view. Moreover, the QPO period $P \simeq 54$~ks is different from that
detected in the previous regular QPE phase ($\simeq 32$~ks), so that
previous QPOs do not enhance, at least formally, the QPO significance
in XMM12. On the other hand, the very fact that a QPO was consistently
detected in all \xmm\ exposures during the previous regular QPE phase
\citep{2023A&A...670A..93M} suggests the need to explore the QPO candidate in
XMM12 further.

\section{Properties of the new QPE phase}
\label{sec:newQPEprop}

The data so far suggest that QPEs and QPOs are linked and could
represent different aspects of the same phenomenon
\citep[see][Sect.~4]{2023A&A...670A..93M}. We cannot exclude that QPOs
represent secondary, longer-lived (broader) QPEs with duty cycles
$\sim 100$\% that follow the primary ones with a delay of
$8$-$10$~ks. Moreover, the QPO (or secondary QPE) is only present
following strong enough QPEs (see Appendix~\ref{sec:Aratio}). During the
XMM12 observation, the weak QPE has roughly the same intensity as the
irregular ones in XMM6 that were not associated with a QPO (see
e.g. Fig.~\ref{fig:QPEsT4} or \ref{fig:QPEsRatios}). We then speculate
that the period of the QPO candidate in XMM12 is representative of the
typical separation between strong QPEs only, $T_{\rm sum}^{\rm (new)}
= T_{\rm long}^{\rm (new)}+T_{\rm short}^{\rm (new)}$, rather than
between consecutive QPEs, so that $T_{\rm sum}^{\rm (new)} \simeq P =
(54\pm 4)$~ks in the new QPE phase.

\begin{figure}
\centering \includegraphics[width=0.96\columnwidth]{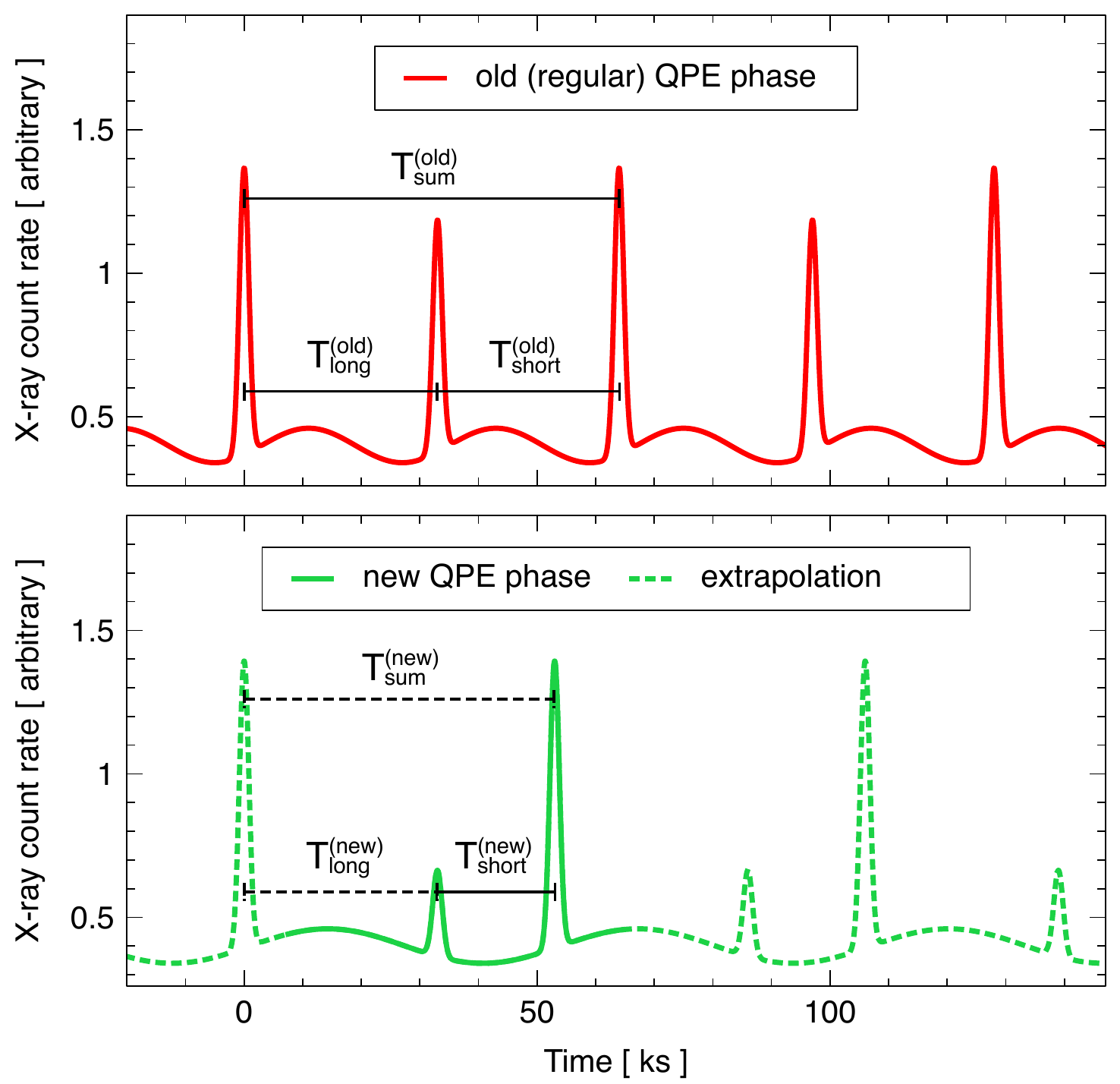}
\caption{Comparison between old and new QPE phases. Upper panel:
  Typical model light curve for observations during the old
  regular QPE phase. Lower panel: Qualitative representation of the
  light curve from the XMM12 observation (new QPE phase; solid line)
  and one possible extrapolation of longer-term behaviour based on the
  arguments discussed in Sect.~\ref{sec:newQPEprop} and Appendix~\ref{sec:Aratio}
  (dashed line). The light curves were normalised so that the
  intensity of the strong QPEs is the same in both panels. The same
  quiescent level is assumed for visual clarity.}
\label{fig:oldnewQPEphases}
\end{figure}

There is another independent way to estimate the recurrence time
$T_{\rm sum}^{\rm (new)}$ between QPEs of the same type during
XMM12. As discussed in Appendix~\ref{sec:Aratio}, the correlation
between the intensity ratio of consecutive QPEs and the recurrence
time between them can be used to infer $T_{\rm sum}^{\rm (new)} =
(52\pm 4)$~ks. This is fully consistent with the estimate obtained
from the QPO-like variability and the two independent estimates
overlap for $T_{\rm sum}^{\rm (new)} \simeq 53\pm 3$~ks. Then, by
definition, one has $T_{\rm short}^{\rm (new)} \simeq 20$~ks and
$T_{\rm long}^{\rm (new)} \gtrsim 27$~ks (and, possibly, $\simeq
33$~ks) in the new QPE phase, to be compared with $T_{\rm short}^{\rm
  (old)} \simeq 31$~ks and $T_{\rm long}^{\rm (old)} \simeq 33$~ks
summing up to $T_{\rm sum}^{\rm (old)} \simeq 64$~ks in the old
(regular) QPE phase. The recurrence time fluctuation $\Delta T_{\rm
  rel} = \left(T_{\rm long}-T_{\rm short}\right)/<T_{\rm rec}>$ (where
$<T_{\rm rec}> = T_{\rm sum}/2$), then increased from $\simeq 6$\% in
the old phase to $\gtrsim 30$\% in the new
one. Figure~\ref{fig:oldnewQPEphases} shows a comparison between old
and new QPE phases together with a possible extrapolation of the new
phase to longer timescales based on the (somewhat speculative)
arguments presented above.

\section{Quiescent luminosity threshold for QPEs}
\label{sec:Lthreshld}

\begin{figure}
\centering \includegraphics[width=0.96\columnwidth]{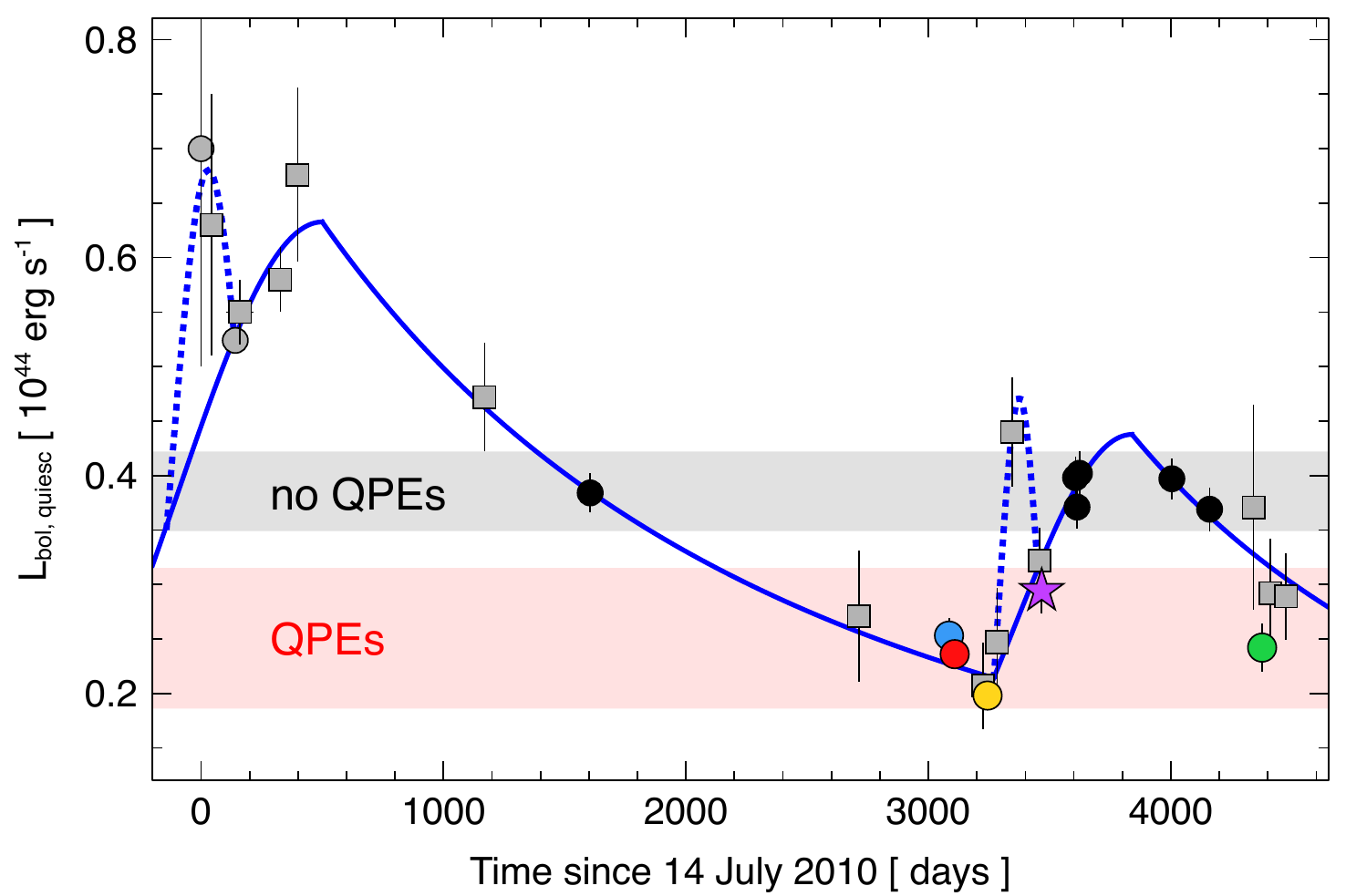}
\caption{Quiescent luminosity long-term evolution. Shown is the $L_{\rm
    bol}$ evolution of the quiescent emission over the past $\sim
  12$~yr. The dotted-solid line is a possible model discussed in
  \citet{2023A&A...670A..93M}. The grey data points refer to observations
  that are too short to ensure the detection of QPEs (squares for \swift\ and circles
  for \xmm\ data). Coloured and black data points represent instead long
  enough observations respectively with and without QPEs. The purple star denotes the
  XMM6 observation with irregular QPEs. A \cha\ observation
  (exhibiting three QPEs) performed between the XMM4 and XMM5 observations was omitted as the
  corresponding quiescent luminosity is highly uncertain (see
  Table~\ref{tab:obs}).}
\label{fig:QPEthreshold}
\end{figure}

To place all QPE observations in context, we updated the quiescent
$L_{\rm bol}$ long-term light curve of GSN~069 in
Fig.~\ref{fig:QPEthreshold} (see Table~\ref{tab:obs}). This is the same
as Fig.~20 in \citet{2023A&A...670A..93M}, with the addition of the
XMM12 observation and of one data point from the latest {\it Neil
  Gehrels} \swift\ monitoring (actually comprising a few short
exposures combined). Figure~\ref{fig:QPEthreshold} shows that QPEs are not found in
the highest luminosity observations, and are instead observed only below a
quiescent bolometric luminosity threshold $L_{\rm thresh} \simeq
3\times 10^{43}$~erg~s$^{-1}$. Assuming, for ease of scaling, a black
hole mass of $M_{\rm BH} = 10^6$~M$_\sun$, and considering the
systematic uncertainties on the estimated bolometric luminosity
discussed in \citet{2023A&A...670A..93M}, the luminosity threshold
translates into an Eddington ratio threshold of $\lambda_{\rm thresh}
\simeq 0.4\pm 0.2$. We note that we define $\lambda_{\rm thresh}$ as
that below which QPEs are present. On the other hand, QPEs are absent
above a slightly higher threshold (a factor of $\simeq 1.3$) because
of the luminosity gap with no observations (see
Fig.~\ref{fig:QPEthreshold}).

In order to study the QPE spectral properties, we first assumed
that QPEs are an additional component with respect to disc emission,
and we extracted intrinsic quiescence-subtracted X-ray spectra of all
QPE peaks, modelling them with a simple absorbed black body (see
Appendix~\ref{sec:spectral}). Figure~\ref{fig:QPEsT4} shows the bolometric
QPE luminosity (at peak) as a function of rest-frame temperature. The
regular QPEs (XMM3 to XMM5) peak at $3.7$-$5.2\times
10^{42}$~erg~s$^{-1}$ with a typical temperature of $90$-$100$~eV
(with a colder outlier during XMM5). The irregular QPEs (XMM6) are
significantly less luminous (and colder) than the regular ones. The
strong and weak QPEs in the new phase (XMM12) are remarkably
consistent with the previously detected regular and irregular QPEs
respectively.

The data are best described by an $L_{\rm bol}\propto T^q$ relation
with $q=3.1 \pm 0.2$. However, as is clear from Fig.~\ref{fig:QPEsT4}, a
$q=4$ solution (dotted lines) corresponding to constant-area black-body
emission cannot be excluded with high significance. Assuming black-body
emission and an emitting area $A=4\pi R^2$, QPEs at peak are
consistent with an emitting region with $R_{\rm peak}\simeq
5$-$10\times 10^{10}$~cm, where we ignore potentially relevant
scattering effects \citep{2021MNRAS.507L..24M}. We also point out
that, since QPEs only carry $\sim 10$-$20$\% of the overall bolometric
luminosity \citep{2019Natur.573..381M,2023A&A...670A..93M}, they
become undetectable once their intrinsic temperature is of the order
of that of the quiescent emission. The lack of QPEs above
$\lambda_{\rm thresh}$ can therefore be taken as an indication that
the contrast between QPE peak and quiescent temperatures is
intrinsically close to unity above threshold, making it impossible to
unambiguously identify QPEs against the quiescent emission. As
discussed in Appendix~\ref{sec:lackQPEs}, the lack of QPEs above
threshold is most likely intrinsic, and not just an effect of the
brighter (and hotter) quiescent level. On the other hand, if QPEs are
not an extra component superimposed on the disc emission, they must
originate from an area $\sim 6$-$30$ times smaller than that
associated with the quiescent state, the smallest regions being
associated with the hottest, most luminous QPEs (see
Appendix~\ref{sec:spectral}).

\begin{figure}
\centering \includegraphics[width=0.96\columnwidth]{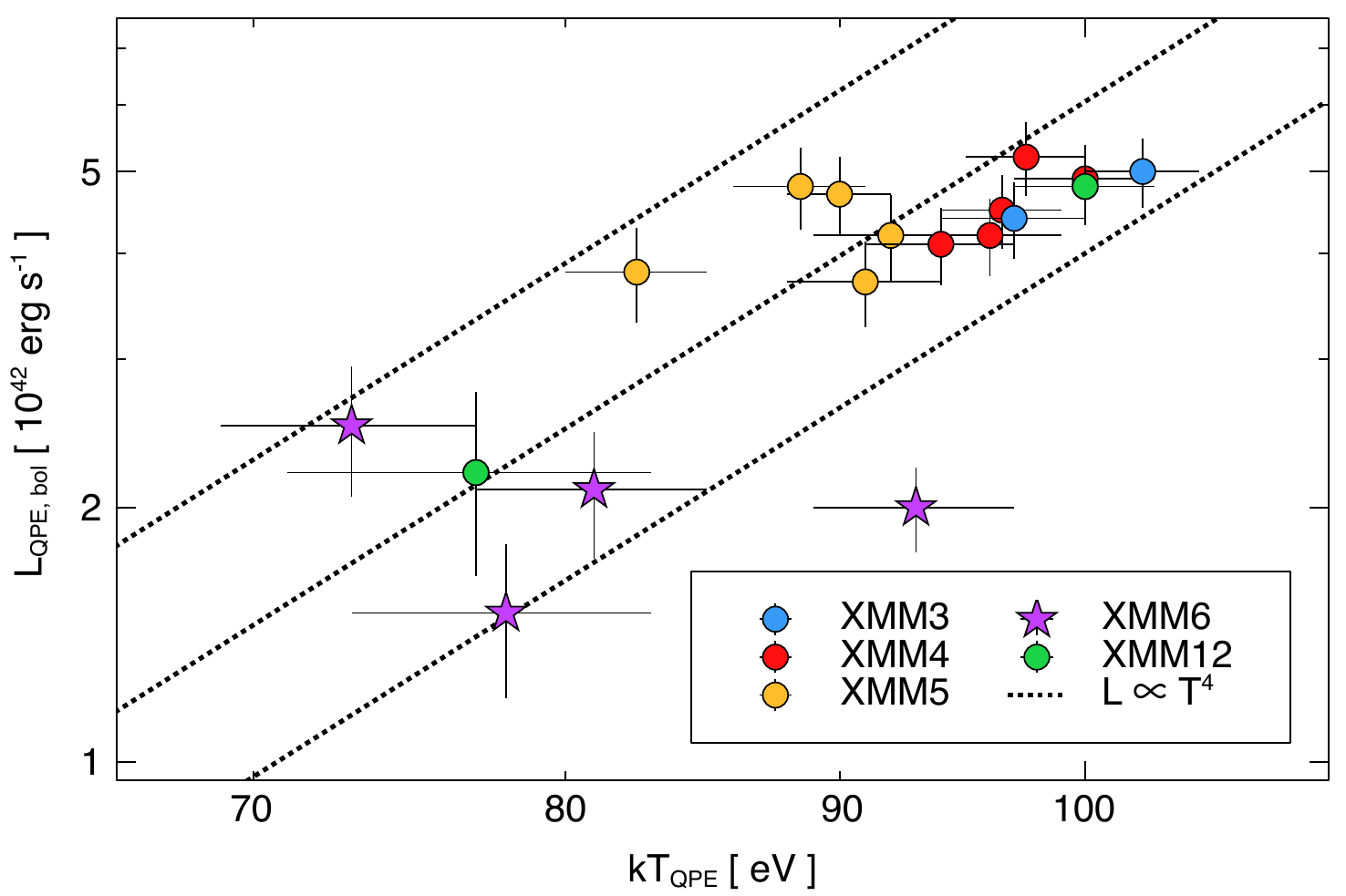}
\caption{QPE $L_{\rm bol}$-$T$ relation. The QPE-only $L_{\rm
    bol}$ is shown as a function of QPE rest-frame temperature from
  quiescence-subtracted QPE spectra assuming a simple absorbed
  black-body model. The dotted lines show $L_{\rm bol} \propto T^4$
  relations with different normalisation. The actual best-fitting
  relation is somewhat shallower with index $\simeq 3.1$.}
\label{fig:QPEsT4}
\end{figure}

\section{Discussion}
\label{sec:discussion}

Several theoretical models have been presented to explain the physical
origin of the QPE phenomenon. They are based on different flavours of
disc instabilities
\citep{2021ApJ...909...82R,2022ApJ...928L..18P,2023arXiv230502071P,2022arXiv221100704K,2023A&A...672A..19S},
impacts between an orbiting companion in an extreme mass-ratio
inspiral (EMRI) system and the existing accretion flow about the
primary
\citep{2021ApJ...917...43S,2021ApJ...921L..32X,2023arXiv230316231L,2023arXiv230400775F,2023arXiv230403670T},
or episodic mass transfer at pericentre from stars or white dwarfs in
a variety of configurations
\citep{2020MNRAS.493L.120K,2022MNRAS.515.4344K,2023MNRAS.520L..63K,2022ApJ...930..122C,2022ApJ...933..225W,2022A&A...661A..55Z,2022ApJ...926..101M,2022arXiv221008023L,2022ApJ...941...24K,2023ApJ...945...86L}. Our
observational results present new challenges for all theoretical
models proposed to date, and we briefly discuss some of their
implications below.

\emph{Recurrence time variation.} Models invoking mass transfer at
pericentre as well as the majority of disc instability models predict
strictly periodic behaviour for consecutive QPEs with $T_{\rm long}
\simeq T_{\rm short}$. Some fluctuation of the recurrence time is
expected, but the observed $\Delta T_{\rm rel} \gtrsim 30$\% in the
new phase (and its factor of $\gtrsim 5$ increase with respect to the
previous phase) is likely too large to be accounted for naturally
\citep{2022MNRAS.515.4344K,2022ApJ...941...24K,2023ApJ...945...86L}. In
the case of disc-secondary impacts, where two QPEs per orbit are
produced, the difference between $T_{\rm long}$ and $T_{\rm short}$ in
the previous regular phase could be explained by assuming a nearly
circular orbit with eccentricity $e\simeq 0.05$-$0.1$. Hence, within
the impacts model, the $\Delta T_{\rm rel}$ variation could be related
to an increase of the EMRI orbital eccentricity between the old and
new QPE phases (from $e\simeq 0.05$-$0.1$. to $e\gtrsim 0.3$). This
could be achieved, for instance, by interaction with a third body
through the Zeipel-Lidov-Kozai mechanism \citep[see the review
  by][]{2016ARA&A..54..441N}. In GSN~069, the third body might
actually be an evolved star on a $\simeq 9$~yr period orbit that was
partially disrupted twice at pericentre giving rise to TDE~1 and TDE~2
\citep{2023A&A...670A..93M}. If the secondary EMRI component is a
star, eccentricity (and period) variation may also arise due to 
extreme mass-transfer episodes (perhaps inducing TDE~2) between the two
QPE phases \citep{2013ApJ...771L..28M,2015MNRAS.449..771G}, as
discussed by \citet{2023arXiv230316231L}.

\emph{Quiescence Eddington ratio threshold for QPE appearance.} Models
in which QPEs represent the fast transient evolution of the disc
emission naturally depend on disc properties and might therefore be
made consistent with the quiescent luminosity threshold. These include
disc instability models \citep[see e.g.][]{2021ApJ...909...82R,2023A&A...672A..19S,2023arXiv230502071P},
burst of mass accretion rate limited to the very innermost regions
\citep[possibly consistent with mass transfer from a highly eccentric
  orbit;
  see][]{2020MNRAS.493L.120K,2022MNRAS.515.4344K,2022ApJ...930..122C,2022ApJ...933..225W},
or the transient formation of an inner, optically thick warm corona
\citep{2019Natur.573..381M}. Considering disc instabilities as
representative examples, the disc may stabilise at high mass accretion
rates or the instability period may increase to the extent that QPEs are
undetected during typical \xmm\ exposures
\citep[see][]{2023arXiv230502071P}.

On the other hand, models in which QPEs are an extra emission
component superimposed on that from the disc appear to be viable
only if QPEs are produced by an interaction with the existing flow as
independent additive QPEs with typical properties (luminosity and
temperature) would have been easily detected even above threshold (see
Appendix~\ref{sec:lackQPEs}). Additive QPE models in which QPEs
originate from an interaction with the accretion flow include, for
instance, disc-secondary impacts
\citep{2021ApJ...921L..32X,2023arXiv230316231L,2023arXiv230400775F,2023arXiv230403670T},
and shocks between the incoming streams and the disc in mass transfer
scenarios \citep{2022arXiv221008023L}. We have also shown that the
lack of QPEs at high luminosities is most likely associated with a
QPE-to-quiescence temperature ratio $kT_{\rm QPE} / kT_{\rm
  quiesc}\sim 1$ above $\lambda_{\rm thresh}$. This appears to be
qualitatively consistent with the work of \citet{2022arXiv221008023L}
and \citet{2023arXiv230400775F}, who show that the temperature ratio
decreases with increasing accretion rate, asymptotically reaching
unity in the latter model. As a final remark, we note that the
luminosity threshold could be artificial, and QPEs might instead only
appear with a certain delay with respect to the TDE peaks (see
Fig.~\ref{fig:QPEthreshold})

\section{Conclusions}
\label{sec:conclusions}

A new \xmm\ observation in July 2022 shows that QPEs have reappeared
in GSN~069 after $\sim 2$~yr of absence. We detect one weak and one
strong QPE separated by $\simeq 20$~ks, a significantly shorter 
recurrence time than ever observed before. The lack of QPEs during the
first $\simeq 27$~ks of the \xmm\ exposures implies that QPEs are no longer 
strictly quasi-periodic or that, at least, $T_{\rm rec}$
fluctuations are now $\Delta T_{\rm rel} \gtrsim 30$\%, a factor of
$\gtrsim 5$ larger than in the previous phase. Under the strong
assumption that the new QPE phase behaves similarly to the previous
regular one, we suggest that the recurrence time between consecutive
QPEs of the same strong or weak type has decreased from $T_{\rm
  sum}^{\rm (old)} \simeq 64$~ks to $T_{\rm sum}^{\rm (old)} \simeq
53$~ks, a speculation that can be tested with future long X-ray
observations. The intensity and temperature of the two consecutive
QPEs in the new phase are remarkably different, unlike at any epoch
observed previously.

Assuming QPE to represent an extra component superimposed on disc
emission, the analysis of all quiescence-subtracted QPE spectra,
together with previous results \citep{2023A&A...670A..93M}, shows that
QPEs are consistent with a region that expands by a factor of $2$-$3$
during the individual QPE evolution, reaching a radius $R_{\rm
  peak}\simeq 5$-$10\times 10^{10}$~cm at peak (ignoring
scattering). Further expansion decreases the contrast between QPE and
quiescence temperatures and leads to the individual QPE decay. If QPEs
are instead associated with the fast evolution of the accretion flow
itself, they must originate from a region with area $\sim 6$-$30$
times smaller than that associated with the quiescent state, the
hottest (and most luminous) QPEs being associated with the smallest
emitting regions.

We also show that QPEs only appear below an Eddington ratio threshold
$\lambda_{\rm thresh} \simeq 0.4\pm 0.2$ of the quiescent emission,
where we assumed a $10^6$~M$_\sun$ for ease of scaling. If the
threshold is confirmed by future observations, QPE models where the
disc plays an important role appear to be particularly
appealing. Models should account for the QPE disappearance above
threshold and we show that this is likely induced by a
QPE-to-quiescence temperature ratio $kT_{\rm QPE}/ kT_{\rm quiesc}
\simeq 1$ above $\lambda_{\rm thresh}$. Theoretical models for the
origin of QPEs face the challenge of explaining self-consistently the
large variation in $\Delta T_{\rm rel}$ from the old to the new phase
and the associated short recurrence time between consecutive QPEs in
XMM12, as well as the presence of two consecutive QPEs with very
different intensities and temperatures at the same epoch. Future long
uninterrupted observations with \xmm\ are crucial to constrain
efficiently the properties of the new QPE phase in GSN~069, which will
enable us to set even tighter constraints on QPE models.

\begin{acknowledgements}

We thank Alessia Franchini, Matteo Bonetti, Norbert Schartel, and
Taeho Ryu for thoughtful discussions and suggestions. We also thank
the anonymous referee for constructive criticism which helped us to
improve the clarity of our presentation. This work is based on
observations obtained with XMM-Newton, an ESA science mission with
instruments and contributions directly funded by ESA Member States and
NASA. We are grateful to the \xmm\ SOC for performing the XMM12
observation as unanticipated TOO. This work made also use of data
supplied by the UK Swift Science Data Centre at the University of
Leicester. MG is supported by the ``Programa de Atracci\'on de
Talento'' of the Comunidad de Madrid, grant number
2018-T1/TIC-11733. RA received support by NASA through the NASA
Einstein Fellowship grant No HF2-51499 awarded by the Space Telescope
Science Institute, which is operated by the Association of
Universities for Research in Astronomy, Inc., for NASA, under contract
NAS5-26555.

\end{acknowledgements}

\bibliographystyle{aa}
\bibliography{biblio}

\newcommand{\noop}[1]{}
\begin{thebibliography}{42}
\expandafter\ifx\csname natexlab\endcsname\relax\def\natexlab#1{#1}\fi

\bibitem[{{Arcavi} {et~al.}(2014){Arcavi}, {Gal-Yam}, {Sullivan}, {Pan},
  {Cenko}, {Horesh}, {Ofek}, {De Cia}, {Yan}, {Yang}, {Howell}, {Tal},
  {Kulkarni}, {Tendulkar}, {Tang}, {Xu}, {Sternberg}, {Cohen}, {Bloom},
  {Nugent}, {Kasliwal}, {Perley}, {Quimby}, {Miller}, {Theissen}, \&
  {Laher}}]{2014ApJ...793...38A}
{Arcavi}, I., {Gal-Yam}, A., {Sullivan}, M., {et~al.} 2014, \apj, 793, 38

\bibitem[{{Arcodia} {et~al.}(2021){Arcodia}, {Merloni}, {Nandra}, {Buchner},
  {Salvato}, {Pasham}, {Remillard}, {Comparat}, {Lamer}, {Ponti}, {Malyali},
  {Wolf}, {Arzoumanian}, {Bogensberger}, {Buckley}, {Gendreau}, {Gromadzki},
  {Kara}, {Krumpe}, {Markwardt}, {Ramos-Ceja}, {Rau}, {Schramm}, \&
  {Schwope}}]{2021Natur.592..704A}
{Arcodia}, R., {Merloni}, A., {Nandra}, K., {et~al.} 2021, \nat, 592, 704

\bibitem[{{Arcodia} {et~al.}(2022){Arcodia}, {Miniutti}, {Ponti}, {Buchner},
  {Giustini}, {Merloni}, {Nandra}, {Vincentelli}, {Kara}, {Salvato}, \&
  {Pasham}}]{2022A&A...662A..49A}
{Arcodia}, R., {Miniutti}, G., {Ponti}, G., {et~al.} 2022, \aap, 662, A49

\bibitem[{{Chakraborty} {et~al.}(2021){Chakraborty}, {Kara}, {Masterson},
  {Giustini}, {Miniutti}, \& {Saxton}}]{2021ApJ...921L..40C}
{Chakraborty}, J., {Kara}, E., {Masterson}, M., {et~al.} 2021, \apjl, 921, L40

\bibitem[{{Chen} {et~al.}(2022){Chen}, {Qiu}, {Li}, \&
  {Liu}}]{2022ApJ...930..122C}
{Chen}, X., {Qiu}, Y., {Li}, S., \& {Liu}, F.~K. 2022, \apj, 930, 122

\bibitem[{{Franchini} {et~al.}(2023){Franchini}, {Bonetti}, {Lupi}, {Miniutti},
  {Bortolas}, {Giustini}, {Dotti}, {Sesana}, {Arcodia}, \&
  {Ryu}}]{2023arXiv230400775F}
{Franchini}, A., {Bonetti}, M., {Lupi}, A., {et~al.} 2023, arXiv e-prints,
  arXiv:2304.00775

\bibitem[{{French} {et~al.}(2020){French}, {Wevers}, {Law-Smith}, {Graur}, \&
  {Zabludoff}}]{2020SSRv..216...32F}
{French}, K.~D., {Wevers}, T., {Law-Smith}, J., {Graur}, O., \& {Zabludoff},
  A.~I. 2020, \ssr, 216, 32

\bibitem[{{Gafton} {et~al.}(2015){Gafton}, {Tejeda}, {Guillochon}, {Korobkin},
  \& {Rosswog}}]{2015MNRAS.449..771G}
{Gafton}, E., {Tejeda}, E., {Guillochon}, J., {Korobkin}, O., \& {Rosswog}, S.
  2015, \mnras, 449, 771

\bibitem[{{Giustini} {et~al.}(2020){Giustini}, {Miniutti}, \&
  {Saxton}}]{2020A&A...636L...2G}
{Giustini}, M., {Miniutti}, G., \& {Saxton}, R.~D. 2020, \aap, 636, L2

\bibitem[{{HI4PI Collaboration} {et~al.}(2016){HI4PI Collaboration}, {Ben
  Bekhti}, {Fl{\"o}er}, {Keller}, {Kerp}, {Lenz}, {Winkel}, {Bailin},
  {Calabretta}, {Dedes}, {Ford}, {Gibson}, {Haud}, {Janowiecki}, {Kalberla},
  {Lockman}, {McClure-Griffiths}, {Murphy}, {Nakanishi}, {Pisano}, \&
  {Staveley-Smith}}]{2016A&A...594A.116H}
{HI4PI Collaboration}, {Ben Bekhti}, N., {Fl{\"o}er}, L., {et~al.} 2016, \aap,
  594, A116

\bibitem[{{Kaur} {et~al.}(2022){Kaur}, {Stone}, \&
  {Gilbaum}}]{2022arXiv221100704K}
{Kaur}, K., {Stone}, N.~C., \& {Gilbaum}, S. 2022, arXiv e-prints,
  arXiv:2211.00704

\bibitem[{{King}(2020)}]{2020MNRAS.493L.120K}
{King}, A. 2020, \mnras, 493, L120

\bibitem[{{King}(2022)}]{2022MNRAS.515.4344K}
{King}, A. 2022, \mnras, 515, 4344

\bibitem[{{King}(2023{\natexlab{a}})}]{2023MNRAS.520L..63K}
{King}, A. 2023{\natexlab{a}}, \mnras, 520, L63

\bibitem[{{King}(2023{\natexlab{b}})}]{2023MNRAS.tmpL..51K}
{King}, A. 2023{\natexlab{b}}, \mnras [\eprint[arXiv]{2303.16185}]

\bibitem[{{Krolik} \& {Linial}(2022)}]{2022ApJ...941...24K}
{Krolik}, J.~H. \& {Linial}, I. 2022, \apj, 941, 24

\bibitem[{{Lin} {et~al.}(2013){Lin}, {Irwin}, {Godet}, {Webb}, \&
  {Barret}}]{2013ApJ...776L..10L}
{Lin}, D., {Irwin}, J.~A., {Godet}, O., {Webb}, N.~A., \& {Barret}, D. 2013,
  \apjl, 776, L10

\bibitem[{{Linial} \& {Metzger}(2023)}]{2023arXiv230316231L}
{Linial}, I. \& {Metzger}, B.~D. 2023, arXiv e-prints, arXiv:2303.16231

\bibitem[{{Linial} \& {Sari}(2023)}]{2023ApJ...945...86L}
{Linial}, I. \& {Sari}, R. 2023, \apj, 945, 86

\bibitem[{{Lu} \& {Quataert}(2022)}]{2022arXiv221008023L}
{Lu}, W. \& {Quataert}, E. 2022, arXiv e-prints, arXiv:2210.08023

\bibitem[{{Manukian} {et~al.}(2013){Manukian}, {Guillochon}, {Ramirez-Ruiz}, \&
  {O'Leary}}]{2013ApJ...771L..28M}
{Manukian}, H., {Guillochon}, J., {Ramirez-Ruiz}, E., \& {O'Leary}, R.~M. 2013,
  \apjl, 771, L28

\bibitem[{{Metzger} {et~al.}(2022){Metzger}, {Stone}, \&
  {Gilbaum}}]{2022ApJ...926..101M}
{Metzger}, B.~D., {Stone}, N.~C., \& {Gilbaum}, S. 2022, \apj, 926, 101

\bibitem[{{Miniutti} {et~al.}(2023){Miniutti}, {Giustini}, {Arcodia}, {Saxton},
  {Read}, {Bianchi}, \& {Alexander}}]{2023A&A...670A..93M}
{Miniutti}, G., {Giustini}, M., {Arcodia}, R., {et~al.} 2023, \aap, 670, A93

\bibitem[{{Miniutti} {et~al.}(2019){Miniutti}, {Saxton}, {Giustini}, {Alexand
  er}, {Fender}, {Heywood}, {Monageng}, {Coriat}, {Tzioumis}, {Read}, {Knigge},
  {Gandhi}, {Pretorius}, \& {Ag{\'\i}s-Gonz{\'a}lez}}]{2019Natur.573..381M}
{Miniutti}, G., {Saxton}, R.~D., {Giustini}, M., {et~al.} 2019, \nat, 573, 381

\bibitem[{{Miniutti} {et~al.}(2013){Miniutti}, {Saxton},
  {Rodr{\'\i}guez-Pascual}, {Read}, {Esquej}, {Colless}, {Dobbie}, \&
  {Spolaor}}]{2013MNRAS.433.1764M}
{Miniutti}, G., {Saxton}, R.~D., {Rodr{\'\i}guez-Pascual}, P.~M., {et~al.}
  2013, \mnras, 433, 1764

\bibitem[{{Mummery}(2021)}]{2021MNRAS.507L..24M}
{Mummery}, A. 2021, \mnras, 507, L24

\bibitem[{{Naoz}(2016)}]{2016ARA&A..54..441N}
{Naoz}, S. 2016, \araa, 54, 441

\bibitem[{{Pan} {et~al.}(2023){Pan}, {Li}, \& {Cao}}]{2023arXiv230502071P}
{Pan}, X., {Li}, S.-L., \& {Cao}, X. 2023, arXiv e-prints, arXiv:2305.02071

\bibitem[{{Pan} {et~al.}(2022){Pan}, {Li}, {Cao}, {Miniutti}, \&
  {Gu}}]{2022ApJ...928L..18P}
{Pan}, X., {Li}, S.-L., {Cao}, X., {Miniutti}, G., \& {Gu}, M. 2022, \apjl,
  928, L18

\bibitem[{{Quintin} {et~al.}(2023){Quintin}, {Webb}, {Guillot}, {Miniutti},
  {Kammoun}, {Giustini}, {Arcodia}, {Soucail}, {Clerc}, {Amato}, \&
  {Markwardt}}]{tormund}
{Quintin}, E., {Webb}, N.~A., {Guillot}, S., {et~al.} 2023, submitted to A\&A

\bibitem[{{Raj} \& {Nixon}(2021)}]{2021ApJ...909...82R}
{Raj}, A. \& {Nixon}, C.~J. 2021, \apj, 909, 82

\bibitem[{{Sheng} {et~al.}(2021){Sheng}, {Wang}, {Ferland}, {Shu}, {Yang},
  {Jiang}, \& {Chen}}]{2021ApJ...920L..25S}
{Sheng}, Z., {Wang}, T., {Ferland}, G., {et~al.} 2021, \apjl, 920, L25

\bibitem[{{Shu} {et~al.}(2018){Shu}, {Wang}, {Dou}, {Jiang}, {Wang}, \&
  {Wang}}]{2018ApJ...857L..16S}
{Shu}, X.~W., {Wang}, S.~S., {Dou}, L.~M., {et~al.} 2018, \apjl, 857, L16

\bibitem[{{{\'S}niegowska} {et~al.}(2023){{\'S}niegowska},
  {Grz{\c{e}}dzielski}, {Czerny}, \& {Janiuk}}]{2023A&A...672A..19S}
{{\'S}niegowska}, M., {Grz{\c{e}}dzielski}, M., {Czerny}, B., \& {Janiuk}, A.
  2023, \aap, 672, A19

\bibitem[{{Sukov{\'a}} {et~al.}(2021){Sukov{\'a}}, {Zaja{\v{c}}ek}, {Witzany},
  \& {Karas}}]{2021ApJ...917...43S}
{Sukov{\'a}}, P., {Zaja{\v{c}}ek}, M., {Witzany}, V., \& {Karas}, V. 2021,
  \apj, 917, 43

\bibitem[{{Tagawa} \& {Haiman}(2023)}]{2023arXiv230403670T}
{Tagawa}, H. \& {Haiman}, Z. 2023, arXiv e-prints, arXiv:2304.03670

\bibitem[{{Terashima} {et~al.}(2012){Terashima}, {Kamizasa}, {Awaki}, {Kubota},
  \& {Ueda}}]{2012ApJ...752..154T}
{Terashima}, Y., {Kamizasa}, N., {Awaki}, H., {Kubota}, A., \& {Ueda}, Y. 2012,
  \apj, 752, 154

\bibitem[{{Wang} {et~al.}(2022){Wang}, {Yin}, {Ma}, \&
  {Wu}}]{2022ApJ...933..225W}
{Wang}, M., {Yin}, J., {Ma}, Y., \& {Wu}, Q. 2022, \apj, 933, 225

\bibitem[{{Webbe} \& {Young}(2023)}]{2023MNRAS.518.3428W}
{Webbe}, R. \& {Young}, A.~J. 2023, \mnras, 518, 3428

\bibitem[{{Wevers} {et~al.}(2022){Wevers}, {Pasham}, {Jalan}, {Rakshit}, \&
  {Arcodia}}]{2022A&A...659L...2W}
{Wevers}, T., {Pasham}, D.~R., {Jalan}, P., {Rakshit}, S., \& {Arcodia}, R.
  2022, \aap, 659, L2

\bibitem[{{Xian} {et~al.}(2021){Xian}, {Zhang}, {Dou}, {He}, \&
  {Shu}}]{2021ApJ...921L..32X}
{Xian}, J., {Zhang}, F., {Dou}, L., {He}, J., \& {Shu}, X. 2021, \apjl, 921,
  L32

\bibitem[{{Zhao} {et~al.}(2022){Zhao}, {Wang}, {Zou}, {Wang}, \&
  {Dai}}]{2022A&A...661A..55Z}
{Zhao}, Z.~Y., {Wang}, Y.~Y., {Zou}, Y.~C., {Wang}, F.~Y., \& {Dai}, Z.~G.
  2022, \aap, 661, A55

\end{thebibliography}

\begin{appendix}

\section{Recurrence time between QPEs of the same strong or weak type}
\label{sec:Aratio}

As discussed by \citet{2023A&A...670A..93M}, a QPO was detected in all
\xmm\ observations during the regular QPE phase when the X-ray light
curves were folded at the average recurrence time between QPEs. In the
original (unfolded) time series, the QPO was seen as excess emission
of the quiescent level with respect to its baseline count rate $\simeq
8$-$10$~ks after most QPEs. No QPO was detected during the irregular
QPE phase (XMM6), when QPEs were significantly weaker. The QPE and QPO
properties therefore suggest that quiescent level excess emission
(inducing the QPO-like variability in folded light curves) is only
seen after strong enough QPEs.

The weak QPE in XMM12 has roughly the same intensity as those during
the irregular, QPO-less XMM6 observation (see Fig.~\ref{fig:QPEsT4} or
~\ref{fig:QPEsRatios}). If QPEs in XMM12 follow the same rules as
during the previous phase, only strong QPEs in
the new phase are expected to give rise to excess emission of the quiescent
level. Hence, the period of the QPO candidate in XMM12 corresponds to
the typical, average recurrence time between strong QPEs only. As long
(short) recurrence times always followed strong (weak) QPEs in the
previous regular phase (see the upper panel of
Fig.~\ref{fig:oldnewQPEphases}), the recurrence time between strong
QPEs is simply $T_{\rm sum}^{\rm (new)} = T_{\rm long}^{\rm (new)} +
T_{\rm short}^{\rm (new)}$ so that $T_{\rm sum}^{\rm (new)} \simeq P =
(54\pm 4)$~ks in XMM12.

\begin{figure}[t!]
\centering \includegraphics[width=0.96\columnwidth]{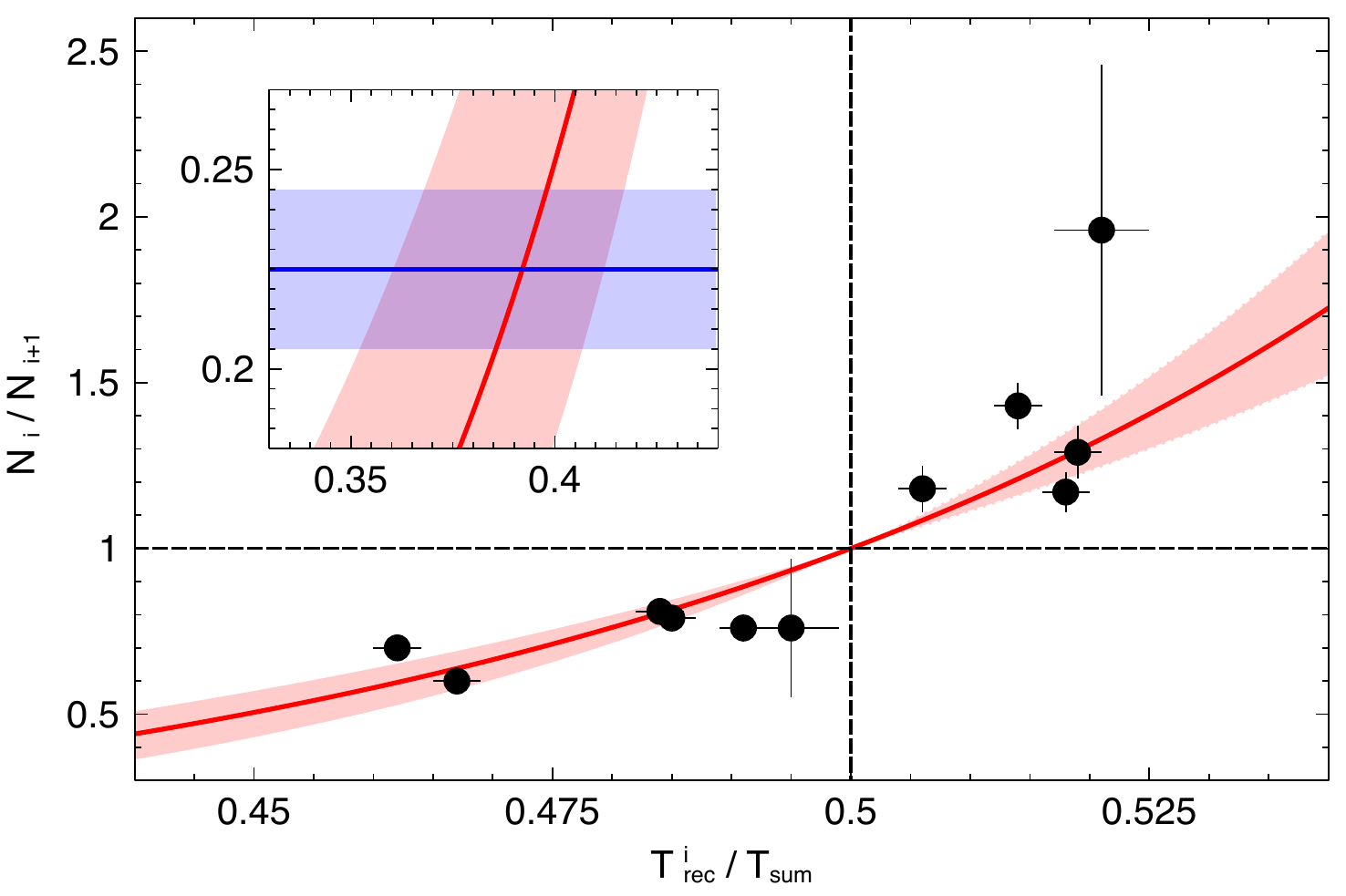}
\caption{QPE intensity-recurrence relation. The ratio between the
  intensity of consecutive QPEs (in the $0.4$-$1$~keV band) is shown
  as a function of the recurrence time between them, which is
  normalised to the average sum of consecutive long and short
  intervals (i.e. to the average separation between QPEs of the same
  type). The vertical dotted line separates short recurrence times
  from long ones, while the horizontal line separates weak-to-strong
  QPE pairs from strong-to-weak pairs. The best-fitting model is shown
  as a solid line and the 1$\sigma$ uncertainty is represented by the shaded
  area. The inset in the upper left quadrant shows the intersection
  between the extrapolated model and the intensity ratio of the
  observed QPEs during XMM12.}
\label{fig:new4}
\end{figure}

A further tool that can be used to estimate $T_{\rm sum}^{\rm (new)}$
comes from the correlation noted by \citet{2023A&A...670A..93M} (see
their Fig.~4) between the intensity ratio of consecutive QPEs in the
previous regular phase and the recurrence time between them. We define
the intensity $N_i$ of QPE$_i$ as the normalisation of the Gaussian
describing it in the $0.4-1$~keV band\footnote{We use the restricted
  $0.4-1$~keV band to include three QPEs from a \cha\ observation
  whose soft X-ray degradation does not allow us to consider data
  below $0.4$~keV.}, $T_{\rm rec}^i$ as the recurrence time
(separation) between QPE$_i$ and the consecutive QPE$_{i+1}$, and
$T_{\rm sum}$ as the average recurrence time between consecutive QPEs
of the same type ($T_{\rm sum} = <T_{\rm rec}^i + T_{\rm rec}^{i\pm
  1}>$). The relationship between consecutive QPE intensity ratios and
recurrence times (as a fraction of $T_{\rm sum}$) is shown in
Fig.~\ref{fig:new4}. This is the same as Fig.~4 in
\citet{2023A&A...670A..93M}, except fot the best-fitting model and inset.

Here we assume that the correlation is valid in the whole interval
$0\leq T_{\rm rec}^i~/T_{\rm sum} \leq 1$ and that $N_i~/N_{i+1} =1$
when $T_{\rm rec}^i~/T_{\rm sum}=0.5$, as suggested by the
data. Hence, we look for a model that satisfies the following
constraints: (i) $N_{i}~/N_{i+1} \to 0$ when $T_{\rm rec}^i~/T_{\rm
  sum} \to 0$; (ii) $N_{i}~/N_{i+1} \to \infty$ when $T_{\rm
  rec}^i~/T_{\rm sum} \to 1$; (iii) $N_i~/N_{i+1} =1$ when $T_{\rm
  rec}^i~/T_{\rm sum}=0.5$. The simplest model that satisfies the
above constraints is $y=\left[ x/\left( 1-x\right)\right]^\alpha$,
where $y=N_i/N_{i+1}$ and $x=T_{\rm rec}^i/T_{\rm sum}$; the
best-fitting model is shown in Fig.~\ref{fig:new4} as a solid line. We
must point an important caveat: the best-fitting model is not
only phenomenological, but is also based on a series of assumptions
driven by the relatively few data points that populate
Fig.~\ref{fig:new4}. We cannot exclude, for example, that had we
detected more QPEs, the two empty quadrants in Fig.~\ref{fig:new4}
would have been populated as well, which would invalidate our
conclusions below.

Although the statistical quality of the fit is poor (reduced
$\chi^2_\nu \simeq 3.5$ with $\alpha = 3.4\pm 0.8$), and with the
caveat expressed above, we can in principle derive $T_{\rm sum}$
from any measure of $N_i/N_{i+1}$ and the corresponding $T_{\rm
  rec}^i$ between them. By extrapolating the best-fitting model to the
observed $N_i/N_{i+1}$ in the $0.4$-$1$~keV band during XMM12 (see
Table~\ref{tab:baseline}), we derive $T_{\rm sum}^{\rm (new)} = (52\pm
4)$~ks. Although this estimate is obtained by extrapolating down to
low $T_{\rm rec}^i/T_{\rm sum}$ a phenomenological model based on data
in a significantly narrower range, the two independent estimates of
$T_{\rm sum}^{\rm (new)}$ derived from the QPO period and from the QPE
intensity ratios ($54\pm 4$~ks and $52\pm 4$~ks) are fully consistent
with each other, which provides some support to the overall arguments.

\section{QPE spectral properties}
\label{sec:spectral}

In order to study the spectral properties of the quiescent emission
and QPEs at all epochs, we extracted EPIC pn quiescent and QPE X-ray
spectra from all \xmm\ observations with QPEs. The former are
extracted from the whole exposures, QPEs excised, while the latter
from time intervals of $~\simeq 1$~ks centred on QPE peaks. There are
two possible ways of interpreting QPE spectra. One can either assume
that QPEs are an extra additive emission component superimposed on the
otherwise stable disc emission, or that they represent the fast
transient evolution of the disc emission itself.\footnote{The former
  scenario is consistent with a series of proposed theoretical models:
  disc-companion collisions, stream-disc and stream-stream
  interactions, or accretion of the mass transferred from an orbiting
  star (provided that the incoming streams do not intercept the disc);
  the last can instead be associated, for example, with disc
  instability models, with scenarios invoking an enhanced mass accretion rate
  in the inner accretion flow during QPEs, or with the transient
  formation of a warm optically thick corona in the innermost disc
  region.} In either case, we are interested in deriving the
properties of QPEs with respect to, and in comparison with, the
quiescent state.

We first considered the additive scenario in which QPEs are superimposed on
the quiescent emission, and we discuss the alternative point of view
at the end of the section. We  constructed QPE intrinsic
(peak) X-ray spectra by subtracting the quiescence from QPEs, as in
\citet{2023A&A...670A..93M}. All spectra were grouped to a minimum of
$20$ counts per energy bin. We fitted jointly all QPE intrinsic
spectra in the $0.3$-$1$~keV band\footnote{Background typically
  dominates above $\sim 1$~keV while uncertain calibration of the EPIC
  pn suggests to discard data below $0.3$~keV.} using $\chi^2$
minimisation in {\tt xspec} and a simple spectral model comprising
Galactic absorption (the {\tt{Tbabs}} model in {\tt{xspec}}) with
$N_{\rm H}$ fixed at $2.3\times 10^{20}$~cm$^{-2}$
\citep{2016A&A...594A.116H}, and a black body (at $z=0.0181$)
representing QPE peak emission, modified by extra intrinsic absorption
(the {\tt{zTbabs}} model in {\tt{xspec}}). The latter is a simplified
version of the warm absorber used by \citet{2023A&A...670A..93M} and
it is chosen because the spectral quality of the QPE spectra prevented
us from constraining the ionisation parameter well. We initially left the
intrinsic absorber column density free to vary between the different
observations. All column densities turned out to be consistent with
each other with the exception of the XMM4 observation where no
intrinsic absorption was preferred. However, by forcing all $N_{\rm H,
  z}$ to be the same, the statistical quality of the fit was basically
unaffected ($\chi^2 /\nu = 666/704$ versus $\chi^2 /\nu = 661/700$),
so that we forced constant, observation-independent $N_{\rm H, z}$ for
simplicity, measuring $N_{\rm H, z} = (2.8\pm 0.8)\times
10^{20}$~cm$^{-2}$.

\begin{figure}
\centering \includegraphics[width=0.96\columnwidth]{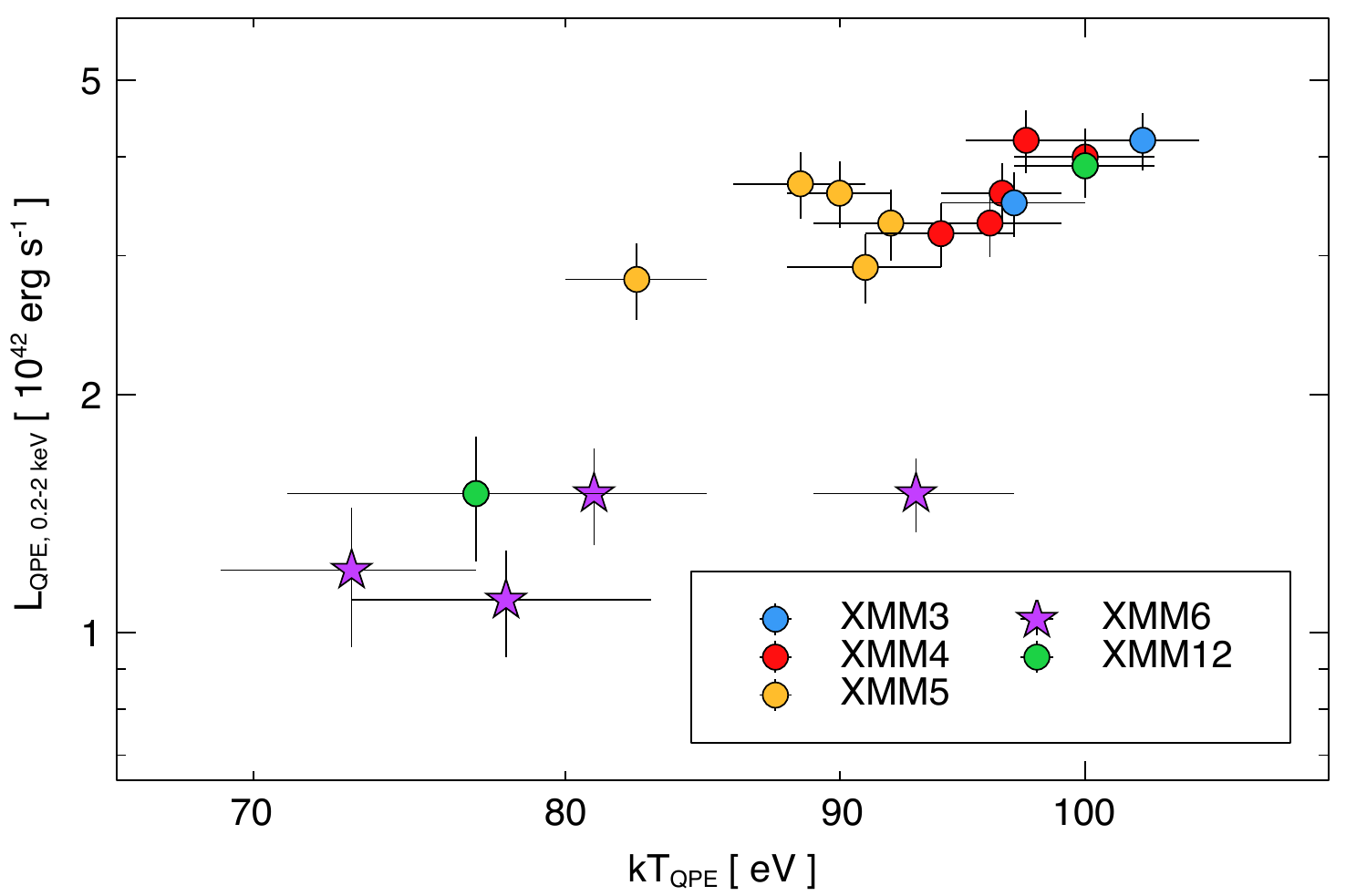}
\caption{QPE-only $L_X$-$T$ relation. The $0.2$-$2$~keV
  X-ray intrinsic peak luminosity of QPEs is shown as a function of rest-frame
  temperature. All spectra are quiescence-subtracted and they are
  described using an absorbed redshifted black-body model.}
\label{fig:QPEsLx}
\end{figure}

In Fig.~\ref{fig:QPEsLx}, we show the $0.2$-$2$~keV X-ray luminosity
of all QPE peaks as a function of their rest-frame temperature. As all
spectra are super-soft, small differences in $N_{\rm H, z}$ (or in the
adopted absorption model) induce relatively large variations in the
derived X-ray luminosities whose absolute values therefore should be
taken with some caution. However, the trend shown in
Fig.~\ref{fig:QPEsLx} is preserved for any reasonable absorption model
and parameters. QPEs during the previous regular phase (XMM3 to XMM5)
as well as the stronger QPE in XMM12 have peak X-ray luminosity in the
range of $2.8$-$4.2\times 10^{42}$~erg~s$^{-1}$ with a typical
temperature of $90$-$100$~eV, while QPEs during the irregular phase
(XMM6) and the weak QPE in XMM12 peak at $1.1$-$1.5\times
10^{42}$~erg~s$^{-1}$ and are characterised by a lower peak
temperature. The $L$-$T$ correlation is shown in terms of the
bolometric (black-body) luminosity in Fig.~\ref{fig:QPEsT4} (see
Sect.~\ref{sec:Lthreshld} for a discussion).

\begin{figure}
\centering \includegraphics[width=0.96\columnwidth]{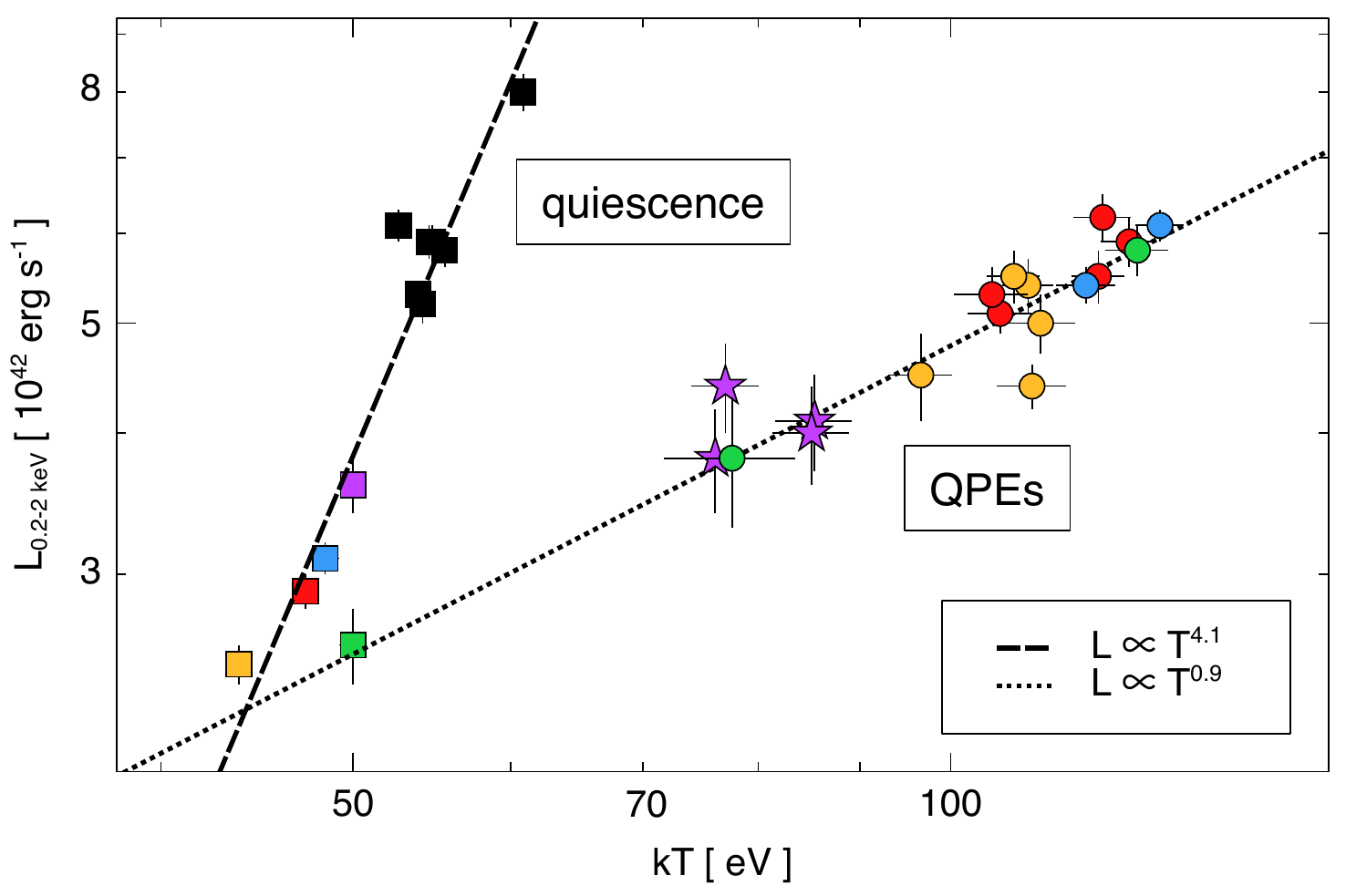}
\caption{Total QPE and quiescent emission $L_X$-$T$
  relation. The $0.2$-$2$~keV X-ray luminosity is shown as a function
  of rest-frame temperature. The squares denote results from fits to
  the quiescent spectra only; the circles are from fits to the QPE
  spectra (not quiescence-subtracted). The same colour scheme as in
  Fig.~\ref{fig:QPEsLx} is adopted, and black data points denote
  observations with no QPEs. All spectra are described using the same
  absorbed redshifted disc model ({\tt diskbb}). The dashed (dotted)
  lines are best-fitting relations for the quiescence (QPEs) of the
  form $L_X \propto T^q$ resulting in $q_{\rm quiesc} \simeq 4.1$ and
  $q_{\rm QPEs}\simeq 0.9$.}
\label{fig:QPEsLx2}
\end{figure}

We also considered the alternative scenario in which QPEs are not an
additive emission component, but rather represent a transient and fast
variation of the disc emission itself (or part of it). If so, the
simplest way of describing QPEs is to consider the whole QPE X-ray
spectrum (that is without subtracting the quiescence), modelling it
with a single spectral component, as done for the quiescence. We then
used the same spectral model for both QPE and quiescent spectra,
namely a multi-temperature disc model ({\tt diskbb} in {\tt
  xspec}). We adopted the same absorption model  (a simple
redshifted neutral absorber), that is we neglected the warm
absorber. All derived quiescent X-ray luminosities are therefore
significantly lower than those presented in
\citet{2023A&A...670A..93M} precisely because of the adopted
absorption model. As mentioned, the relatively low quality of the QPE spectra
($\sim 1$~ks of exposure) prevented us from constraining the warm absorber
properties during QPEs so that we preferred to use the simpler neutral
absorber for both quiescence and QPEs, enabling us to compare the
luminosity of the two states without introducing model-dependent
uncertainties in the comparison. We note that the overall trend
(luminosity versus temperature) discussed below is preserved for any
chosen absorption model.

For clarity, we also point out that the QPE luminosity from
non-quiescence-subtracted spectra is not exactly the sum of the
quiescence and QPE-only spectra (the latter derived from
quiescence-subtracted spectra). This is not surprising since the
spectral model is different and thus extrapolates differently in the
$0.2$-$2$~keV band (data are only fitted in the $0.3$-$1$~keV
band). The QPE temperature is also slightly different because of the
different spectral models describing QPEs and because of the slightly
different spectral shape (spectra being either quiescence-subtracted
or not). By re-fitting the non-quiescence-subtracted QPE
spectra with a two-component model (a disc model for the quiescence
and an additional black-body model for QPEs), we indeed recover the
results for the QPE-only luminosity and temperature, as expected.

The resulting quiescent and QPE X-ray luminosity is shown in
Fig~\ref{fig:QPEsLx2} for both the quiescence and QPEs as a function
of (rest-frame) temperature. We also show best-fitting relations of
the form $L_X \propto T^q$ in both cases, and we derive $q_{\rm
  quiesc} \simeq 4.1$ \citep[as noted by][]{2019Natur.573..381M,
  2023A&A...670A..93M}, and $q_{\rm QPEs} \simeq 0.9$. As QPEs do not
align onto the $L\propto T^{\sim 4}$ relation defined by the quiescent
X-ray emission, QPEs and quiescence cannot be associated with disc
emission from the same region. In other words, as already noted by
\citet{2019Natur.573..381M}, QPEs are not produced by global mass
accretion rate variation. This is also consistent with the lack of
QPEs in the optical and UV and signals that only a very limited inner
region on the disc is responsible for
eruptions. Figure~\ref{fig:QPEsLx2} shows that QPEs must originate
from a region with an area $\sim 6$-$30$ times smaller than that
responsible for the X-ray quiescent emission (if eruptions are
black body-like, as the data indicate). Assuming that the quiescent
X-ray emission region has radius $R_{\rm quiesc} \simeq 20~R_g$ ($R_g
= GM_{\rm BH}/c^2$) this means that QPE emission is confined within
radii of the order of $\sim 3.5$-$8~R_g$. The flat $q_{\rm QPEs}$
implies that the hottest (and most luminous) QPEs are associated with
the smallest emitting regions. A very small QPE emitting region could
in principle be consistent with disc instabilities and a very small
inner unstable region \citep[see e.g.][]{2023A&A...672A..19S,2023arXiv230502071P}, with mass
injection at small radii producing a burst in mass accretion rate
limited to the inner annulii \citep[possibly as
  in][]{2020MNRAS.493L.120K,2022MNRAS.515.4344K,2022ApJ...930..122C,2022ApJ...933..225W}
or with the transient formation of an inner, optically thick, warm
corona inducing the emergence of a transient soft X-ray excess
\citep[as discussed by][]{2019Natur.573..381M}.

\section{Ability to detect QPEs}
\label{sec:lackQPEs}

\begin{figure}
\centering \includegraphics[width=0.96\columnwidth]{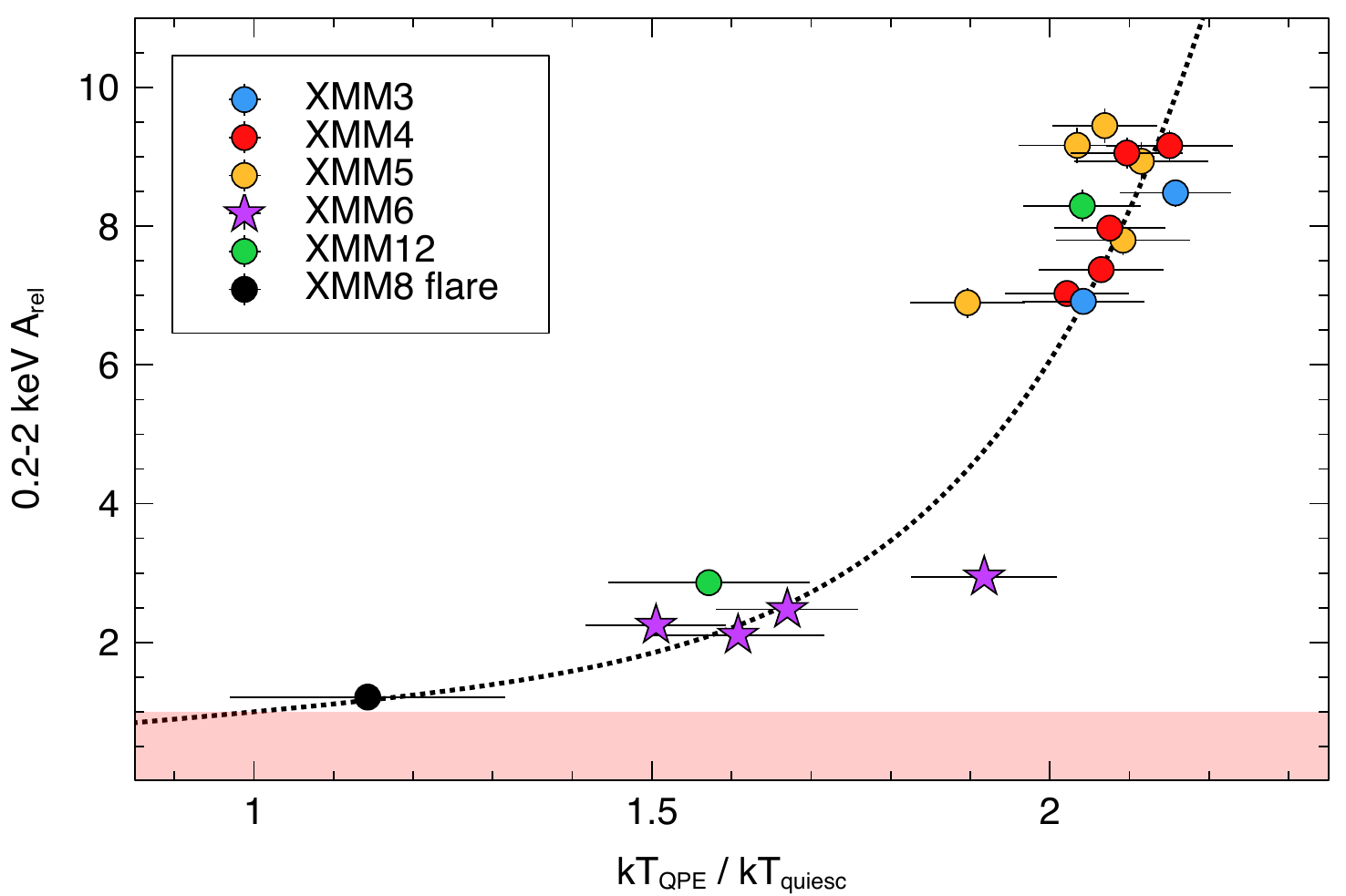}
\caption{Dependence of QPE detection on the eruptions-to-quiescence
  temperature ratio. The relative $0.2$-$2$~keV QPE amplitude is shown
  as a function of the ratio of QPE to quiescent temperatures. The
  horizontal shaded area denotes the region below the QPE
  detection threshold. Also shown (black data point) is the strongest
  X-ray flare seen in any observation with no clearly detected
  QPEs. The dotted line shows a simple fit to the data, whose
  functional form is chosen so that $A_{\rm rel} = 1$ when $kT_{\rm
    QPE}/ kT_{\rm quiesc} =1$ (see text for details).}
\label{fig:QPEsRatios}
\end{figure}

Figures~\ref{fig:QPEsLx} and \ref{fig:QPEsLx2} suggest that QPEs can
be identified efficiently by their different temperatures with respect
to the quiescent level and by the correlation between (peak)
luminosity and temperature. Since QPEs in GSN~069 only carry a small
fraction of the quiescence bolometric luminosity, QPEs with peak
temperature $kT_{\rm QPE}$ close to that of the quiescent emission
$kT_{\rm quiesc}$ cannot be efficiently detected against it as they
would appear as relatively low-amplitude fluctuations of the quiescent
level with no striking energy dependence. In
Fig.~\ref{fig:QPEsRatios}, we show the relative QPE amplitude in the
$0.2$-$2$~keV as a function of the ratio of QPE to quiescent
temperature. We take the QPE temperatures from fits to the
quiescence-subtracted QPE spectra, but we note that
Fig.~\ref{fig:QPEsRatios} is qualitatively very similar when
considering fits to the full QPE spectra (i.e. without subtracting the
quiescence). Hence, our conclusions below are general and do not
depend on whether QPEs are superimposed on the disc emission or
represent instead a transient evolution of the disc emission itself.

The QPE relative amplitude is defined as $A_{\rm rel} = C_{\rm QPE}/
C_{\rm quiesc}$, where $C_{\rm QPE}$ and $C_{\rm quiesc}$ are the
(total) QPE and quiescent $0.2$-$2$~keV count rates respectively. The
relative amplitude is energy-dependent. As an example, the highest
amplitude in the $0.2$-$2$~keV band is $A_{\rm rel, max}\simeq 9.5$,
increasing to $\simeq 82$ in the $0.6$-$2$~keV band, mostly because of
the much lower contribution of the quiescent emission in higher energy
bands. The dotted line in Fig.~\ref{fig:QPEsRatios} shows a
(non-unique) best-fitting model of the form $y = (ax+bx^q)/ (a+b)$,
chosen so that $A_{\rm rel} = 1$ when $kT_{\rm QPE}/ kT_{\rm quiesc}
=1$. The shaded area in Fig.~\ref{fig:QPEsRatios} denotes the region
below the QPE detection threshold.

\begin{figure}
\centering \includegraphics[width=0.96\columnwidth]{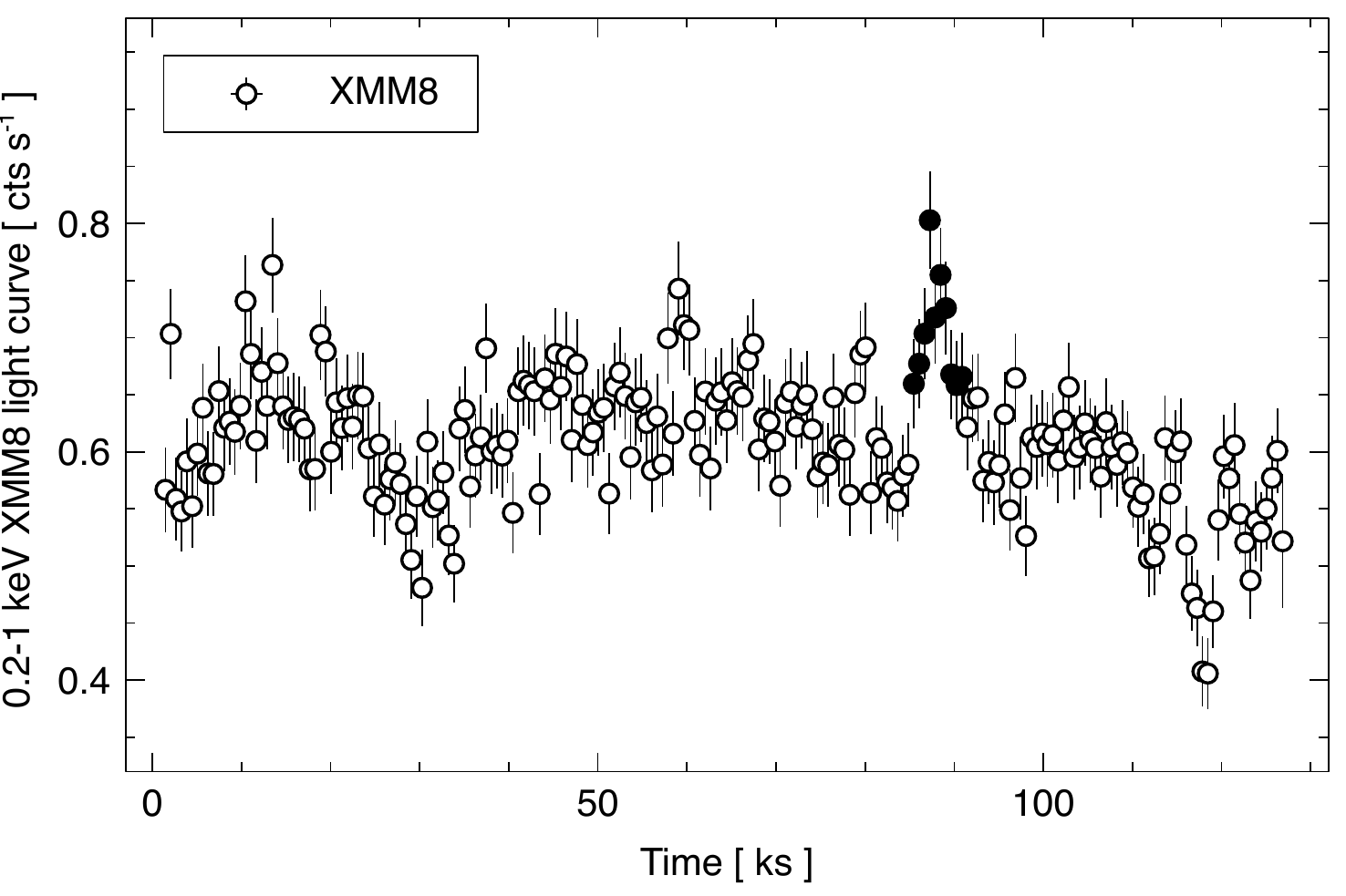}
\caption{Best QPE candidate in QPE-less observations. We show the
  $0.2$-$1$~keV light curve from the XMM8 observation. The most
  prominent flare, that is the best QPE candidate from all
  high-luminosity observations with no unambiguous QPEs, is
  highlighted. We used here a time bin of $600$~s. The flare spectrum
  used to derived the quantities in Fig.~\ref{fig:QPEsRatios} (count
  rate and peak temperature) was extracted from
  a $\sim 1$~ks time-interval around the peak. The empty circles
  define the period during which we accumulated the quiescent
  spectrum.}
\label{fig:xmm8flare}
\end{figure}

The trend in Fig.~\ref{fig:QPEsRatios}, as well as its best-fitting
model, suggest that the detection of QPEs with $kT_{\rm QPE} /kT_{\rm
  quiesc} \lesssim 1.2$ is already challenging in GSN~069. The black
data point in Fig.~\ref{fig:QPEsRatios} refers to the brightest X-ray
flare observed in any of the observations with no clear QPEs, namely
the best QPE candidate from the XMM2 and the XMM7 to XMM11
observations. The light curve from the corresponding observation
(XMM8) is shown in Fig.~\ref{fig:xmm8flare} where the QPE candidate is
highlighted with filled circles. As shown in
Fig.~\ref{fig:QPEsRatios}, we cannot exclude that the XMM8 flare is in
fact a weak QPE with very little temperature contrast with respect to
the quiescent emission.

Hence, a question arises of whether QPEs with similar properties as
those observed during the XMM3 to XMM6 observations could also be
present in the high-luminosity XMM2 and XMM7 to XMM11 observations,
but be undetected against a higher flux and higher temperature
quiescent emission. To answer this question, we selected a set of
representative QPEs, and we studied how they would have appeared in
observations when the quiescent emission was similar to that during
the high-luminosity QPE-less observations. We considered the XMM12
QPEs as representative of QPEs in the previous regular phase (the
strong QPE in XMM12) and of the weaker ones in the irregular phase
(the weak QPE in XMM12), as shown in Fig.~\ref{fig:QPEsRatios}. We
also considered the irregular, weaker QPEs of the XMM6 observation for
completeness. In this case, in order to estimate the appearance of
standard QPEs in observations when the quiescent emission was
brighter, we had to assume the additive nature of QPEs, and we then
proceeded as follows: from the quiescence-subtracted peak QPE spectra,
we recorded the QPE-only $0.2$-$2$~keV count rate $C_{\rm QPE-only}$
and peak temperature $kT_{\rm QPE}$. We then considered the quiescent
spectra of all observations with no QPEs (XMM2 and XMM7 to
XMM11). Since these observations all have similar luminosities and
temperatures \citep[see][Fig.~9]{2023A&A...670A..93M}, we estimated
the average $0.2$-$2$~keV quiescent count rate $C_{\rm quiesc}$ and
temperature $kT_{\rm quiesc}$, and we computed the relative QPE
amplitude $A_{\rm rel} = \left( C_{\rm QPE-only} + C_{\rm
  quiesc}\right)/ C_{\rm quiesc}$ and temperature ratio $kT_{\rm QPE}/
kT_{\rm quiesc}$ for the selected QPEs.

\begin{figure}
\centering \includegraphics[width=0.96\columnwidth]{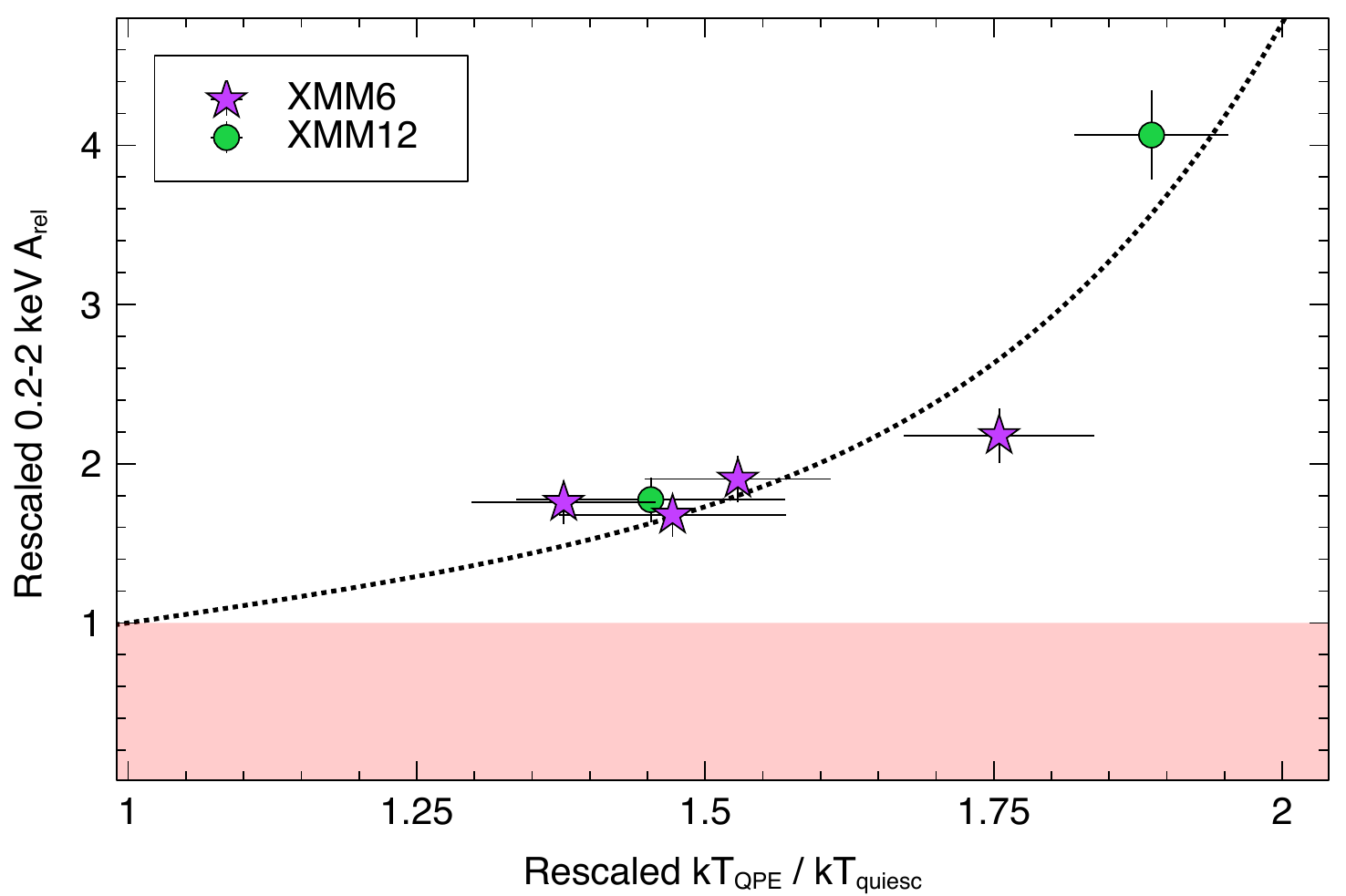}
\caption{Estimated QPE appearance during observations with no clear
  QPEs (XMM2 and XMM7 to XMM11). Shown is the relative $0.2$-$2$~keV
  QPE amplitude as a function of the ratio of QPE to quiescent
  temperatures for the XMM6 and XMM12 observations. $A_{\rm rel}$ and
  $kT_{\rm QPE}/ kT_{\rm quiesc}$ have been rescaled to show how
  additive QPEs would have appeared during high-luminosity
  observations with no unambiguous QPEs (see text for details). The dotted
  line shows a simple fit to the data of the same form as that used
  for Fig.~\ref{fig:QPEsRatios}.}
\label{fig:QPEdetection}
\end{figure}

Figure~\ref{fig:QPEdetection} shows the resulting $A_{\rm rel}$ as a
function of the temperature ratio, which is how these QPEs would have
appeared in observations during which the quiescent level was equal to
the average emission (in both flux and temperature) in
observations with no QPEs, under the assumption that QPEs are an extra
emission component superimposed on the quiescent (disc)
emission. Although both $A_{\rm rel}$ and $kT_{\rm QPE}/ kT_{\rm
  quiesc}$ are slightly reduced due to higher $C_{\rm quiesc}$ and
$kT_{\rm quiesc}$ (compare Fig.~\ref{fig:QPEdetection} with
Fig.~\ref{fig:QPEsRatios}), both the strong QPE in XMM12
(representative of all QPEs during the previous regular phase)
and the weak QPE (as well as all irregular QPEs during XMM6) lie well
above the detection threshold. We can therefore safely conclude that
the lack of clear QPEs in the high quiescent luminosity observations
is not due to the higher flux and temperature of the quiescent
emission, but instead is intrinsic. In other words, if QPEs are present
during these observations (XMM2 and XMM7 to XMM11), they must be
significantly colder than any of the detected QPEs with $kT_{\rm QPE}
/ kT_{\rm quiesc} \lesssim 1$.

\onecolumn

\section{Tables}
\label{sec:tab}

\begin{table}[h!]
        \centering
        \caption{Summary of the \xmm\ and \cha\ pointed observations of GSN~069.}
        \label{tab:obs}
        \begin{tabular}{lcccccc} 
          \hline
\T  & ObsID & Date (start) & Exposure & $L_{\rm bol,\,quiesc}$ & QPEs \B \\
\hline
\T XMM1  & 0657820101 & 2010-12-02 & 13  & $\sim 5.2$ &\textcolor{red}{$\times$} \B \\
\T XMM2  & 0740960101 & 2014-12-05 & 92  & $\sim 3.8$ &\textcolor{red}{$\times$} \B \\
\T XMM3  & 0823680101 & 2018-12-24 & 50  & $\sim 2.5$ &\textcolor{green}{\checkmark}   \B \\
\T XMM4  & 0831790701 & 2019-01-16 & 134 & $\sim 2.4$ &\textcolor{green}{\checkmark}  \B \\
\T Chandra & 22096 & 2019-02-14    & 73  & $\sim 0.09-1.1$&\textcolor{green}{\checkmark} \B \\
\T XMM5  & 0851180401 & 2019-05-31 & 132 & $\sim 2.0$ &\textcolor{green}{\checkmark} \B \\
\T XMM6  & 0864330101 & 2020-01-10 & 131 & $\sim 2.9$ &\textcolor{green}{\checkmark} \B \\
\T XMM7  & 0864330201 & 2020-05-28 & 125 & $\sim 4.0$ &\textcolor{red}{$\times$} \B \\
\T XMM8  & 0864330301 & 2020-06-03 & 126 & $\sim 3.7$ &\textcolor{red}{$\times$} \B \\
\T XMM9  & 0864330401 & 2020-06-13 & 118 & $\sim 4.0$ &\textcolor{red}{$\times$} \B \\
\T XMM10 & 0884970101 & 2021-06-30 & 48  & $\sim 4.0$ &\textcolor{red}{$\times$} \B \\
\T XMM11 & 0884970201 & 2021-12-03 & 45  & $\sim 3.7$ &\textcolor{red}{$\times$} \B \\
\T XMM12 & 0913990201 & 2022-07-07 & 56  & $\sim 2.4$ &\textcolor{green}{\checkmark}  \B\\
          \hline
        \end{tabular}
\tablefoot{The (usable) exposure for all \xmm\ observations (in ks)
  refers to the EPIC-pn camera. We also report the quiescent
  bolometric luminosity (in units of $10^{43}$~erg~s$^{-1}$) for all
  observations. For the XMM1 to XMM11 observations, $L_{\rm
    bol,\,quiesc}$ was taken from \citet{2023A&A...670A..93M}, and we
  used exactly the same procedure and spectral model to estimate it
  during the XMM12 observation. As for the \cha\ observation, we
  report $L_{\rm bol,\,quiesc}$ as estimated by
  \citet{2019Natur.573..381M}, but we point out that $L_{\rm
    bol,\,disc}$ in the \cha\ observation is highly uncertain due to
  the degradation of the \cha\ ACIS detector that forced them to
  ignore data below $0.4$~keV, as well as to the possible appearance
  of a soft X-ray excess component not easily disentangled from the
  disc emission above $0.4$~keV. $L_{\rm bol,\,disc}$ may be
  underestimated by up to a factor of $\sim 2.5$ even in
  \xmm\ observations due to systematic uncertainties on the intrinsic
  optical and UV disc emission \citep[as discussed
    by][]{2023A&A...670A..93M}. The last column indicates whether
  (unambiguous) QPEs were detected during the exposure. }

\end{table}

\begin{table}[h!]
        \centering
        \caption{Baseline best-fitting parameters for the $0.4$-$1$~keV and $0.2$-$1$~keV EPIC-pn light curves from the XMM12 observation.} 
        \label{tab:baseline}
        \begin{tabular}{lccccc} 
          \hline
          \T  QPE type & $T_{\mathrm{rec}}$ & $C$ & $N$  & $\sigma$ & $\chi^2 /\nu$ \B \\
          \hline
          \T  \bf{ 0.4-1~keV} & & &&& \bf{ 480/273} \B \\
          \hline 
          \T QPE$^{\mathrm{(w)}}$ & $19.89\pm 0.05$  & $0.051\pm 0.001$ & $0.25\pm 0.02$ & $557\pm 45$ \B\\
          \T QPE$^{\mathrm{(s)}}$ &              & $-$                & $1.11\pm 0.04$ & $797\pm 21$ \B \\
          \hline
          \T  \bf{0.2-1~keV} & & &&& \bf{ 785/273} \B \\
          \hline 
          \T QPE$^{\mathrm{(w)}}$ & $20.00\pm 0.05$  & $0.339\pm 0.003$ & $0.52\pm 0.05$ & $586\pm 59$ \B\\
          \T QPE$^{\mathrm{(s)}}$ &              & $-$                & $2.00\pm 0.05$ & $885\pm 19$ \B \\
          \hline         
          \end{tabular}
\tablefoot{The baseline model comprises a constant $C$ (in units of
  cts~s$^{-1}$) representing the quiescent level emission, and two
  Gaussian functions with normalisation $N$ (in units of cts~s$^{-1}$)
  and width $\sigma$ (in units of s) describing the two QPEs. $T_{\rm
    rec}$ (in units of ks) is the recurrence time (or time-separation)
  between the two detected QPEs. Superscripts $^{\rm(w)}$ and
  $^{\rm(s)}$ denote the weak and strong QPE respectively. We use a
  time-bin of $200$~s.}
\end{table}

\begin{table}[h!]
        \centering
        \caption{Baseline plus sine function best-fitting parameters for the re-binned $0.2$-$1$~keV EPIC-pn light curves from the XMM12 observation.} 
        \label{tab:sine}
        \begin{tabular}{lcccccc} 
          \hline
          \T  QPE type &$T_{\mathrm{rec}}$ & $C$ & $N$  & $\sigma$ & $A$ & $P$ \B \\
          \hline 
          \T QPE$^{\mathrm{(w)}}$ & $20.05\pm 0.05$  & $0.359\pm 0.007$ & $0.54\pm 0.03$ & $743\pm 48$ & $0.098\pm 0.007$ & $54\pm 4$ \\
          \T QPE$^{\mathrm{(s)}}$ &                & $-$             & $1.87\pm 0.05$ & $913\pm 21$ & $-$ & $-$ \B \\
          \hline         
        \end{tabular}
\tablefoot{The first five columns are the same as in Table~\ref{tab:baseline}. The additional sine function has normalisation $A$ (in units of cts~s$^{-1}$) and period $P$ (in units ks). The best-fitting baseline plus sine function models results in $\chi^2 /\nu = 140/46$. Removing the sine function from the fit produces $\chi^2 /\nu = 550/49$. We use a time-bin of $1\,000$~s.}
\end{table}

\end{appendix}
 
\end{document}